\documentclass[12pt]{article}
\usepackage{amsfonts,amsthm,amsmath,amssymb,graphicx,hyperref}
\usepackage[lmargin=3.0cm, rmargin=1.5cm,tmargin=2.50cm,bmargin=2.50cm]{
geometry}
\usepackage{float}
\usepackage{ulem}
\usepackage[dvips]{color}
\usepackage{colordvi}
\usepackage{url}

\begin{document}

\baselineskip=18pt

\def\gap#1{\vspace{#1 ex}}
\def\be{\begin{equation}}
\def\ee{\end{equation}}
\def\bal{\begin{array}{l}}
\def\ba#1{\begin{array}{#1}}  
\def\ea{\end{array}}
\def\bea{\begin{eqnarray}}
\def\eea{\end{eqnarray}}
\def\beas{\begin{eqnarray*}}
\def\eeas{\end{eqnarray*}}
\def\del{\partial}
\def\eq#1{(\ref{#1})}
\def\fig#1{Fig \ref{#1}} 
\def\re#1{{\bf #1}}
\def\bull{$\bullet$}
\def\nn{\\\nonumber}
\def\ub{\underbar}
\def\nl{\hfill\break}
\def\ni{\noindent}
\def\bibi{\bibitem}
\def\ket{\rangle}
\def\bra{\langle}
\def\vev#1{\langle #1 \rangle} 
\def\lsim{\stackrel{<}{\sim}}
\def\gsim{\stackrel{>}{\sim}}
\def\mattwo#1#2#3#4{\left(
\begin{array}{cc}#1&#2\\#3&#4\end{array}\right)} 
\def\tgen#1{T^{#1}}
\def\half{\frac12}
\def\floor#1{{\lfloor #1 \rfloor}}
\def\ceil#1{{\lceil #1 \rceil}}

\def\mysec#1{\gap1\ni{\bf #1}\gap1}
\def\mycap#1{\begin{quote}{\footnotesize #1}\end{quote}}

\def\lan{\langle}
\def\ran{\rangle}

\def\bit{\begin{item}}
\def\eit{\end{item}}
\def\benu{\begin{enumerate}}
\def\eenu{\end{enumerate}}
\def\a{\alpha}
\def\as{\asymp}
\def\ap{\approx}
\def\b{\beta}
\def\bp{\bar{\partial}}
\def\cA{{\cal{A}}}
\def\cD{{\cal{D}}}
\def\cL{{\cal{L}}}
\def\cP{{\cal{P}}}
\def\cR{{\cal{R}}}
\def\da{\dagger}
\def\de{\delta}
\def\tD{\tilde D}
\def\e{\eta}
\def\ep{\epsilon}
\def\eqv{\equiv}
\def\f{\frac}
\def\g{\gamma}
\def\G{\Gamma}
\def\h{\hat}
\def\hs{\hspace}
\def\i{\iota}
\def\k{\kappa}
\def\lf{\left}
\def\l{\lambda}
\def\la{\leftarrow}
\def\La{\Leftarrow}
\def\Lla{\Longleftarrow}
\def\Lra{\Longrightarrow}
\def\L{\Lambda}
\def\m{\mu}
\def\na{\nabla}
\def\nn{\nonumber\\}
\def\mm{&&\kern-18pt}  
\def\om{\omega}
\def\O{\Omega}
\def\P{\Phi}
\def\pa{\partial}
\def\pr{\prime}
\def\r{\rho}
\def\ra{\rightarrow}
\def\Ra{\Rightarrow}
\def\ri{\right}
\def\s{\sigma}
\def\sq{\sqrt}
\def\S{\Sigma}
\def\si{\simeq}
\def\st{\star}
\def\t{\theta}
\def\ta{\tau}
\def\ti{\tilde}
\def\tm{\times}
\def\tr{\textrm}
\def\Tr{{\rm Tr}}
\def\T{\Theta}
\def\up{\upsilon}
\def\Up{\Upsilon}
\def\v{\varepsilon}
\def\vh{\varpi}
\def\vk{\vec{k}}
\def\vp{\varphi}
\def\vr{\varrho}
\def\vs{\varsigma}
\def\vt{\vartheta}
\def\w{\wedge}
\def\z{\zeta}

\def\psd{\psi^\dagger}

\thispagestyle{empty}
\addtocounter{page}{-1}
\begin{flushright}
TIFR/TH/09-14\\
KEK-TH-1603
\end{flushright}
\begin{center}
\vspace*{0.2cm} {\large \bf Quantum quench in matrix models: Dynamical
  phase transitions,}\\
\vspace*{0.1cm} {\large \bf Selective equilibration and the
  Generalized Gibbs Ensemble} \\
\vspace*{1.0cm} {\bf Gautam~Mandal$^*$ and Takeshi~Morita$^\dagger$}
\vspace*{0.7cm} \\ {\rm $^*$\it Department of Theoretical Physics,
  Tata Institute of Fundamental Research,}\\
\vspace*{0.1cm}
{\it Mumbai 400 005, \rm INDIA}
\vspace*{0.2cm} \\ {$^\dagger$ \it KEK Theory Center, High Energy
  Accelerator Research Organization (KEK)}\\
\vspace*{0.1cm} 
{\it Oho 1-1, Tsukuba, Ibaraki 305-0801, \rm JAPAN.}  
\\ 
\vspace{0.4cm}
{\tt email:
  mandal@theory.tifr.res.in, tmorita@post.kek.jp} \\

\gap2
\today

\vspace*{0.3cm}
{\bf Abstract}

\end{center}

\vspace*{0.3cm}
Quantum quench dynamics is considered in a one dimensional unitary matrix model with a single trace potential.
This model is integrable and has been studied in the context of non-critical string theory.
We find dynamical phase transitions, and study the role of the quantum critical point.
In course of the time evolutions, we find evidence of selective equilibration
for a certain class of observables. The equilibrium is governed by the Generalized Gibbs Ensemble
(GGE) and differs from the standard Gibbs ensemble. We compute the production of entropy which is $O(N)$ for large $N$ matrices. 
An important feature of the equilibration is the appearance of an energy cascade,
reminiscent of the Richardson cascade in turbulence, where we find
flow of energy from initial long wavelength modes to progressively
shorter wavelength excitations. 
We discuss possible implication of the equilibration and of GGE in string theories and higher spin theories.
In another related study, we compute time evolutions in a double trace unitary matrix model, which arises as an effective theory of D2 branes in IIA string theory in the confinement phase.
We find similar equilibrations and dynamical transitions in this matrix model.
The dynamical transitions are related to Gregory-Laflamme transitions in string theory and are potentially connected with the issue of appearance of naked singularities.
\newpage
\tableofcontents

\section{\label{sec:intro}Introduction and Summary}

There has been a lot of progress recently in understanding dynamical
processes in statistical mechanics models as well as in large $N$
gauge theories. An important development in statistical mechanics has
been an extensive study of equilibration in integrable systems
\cite{Polkovnikov:2010yn}. In these studies it is found that
equilibration happens only for a certain class of observables (whose
characterization remains an open problem), and when it does, it is not
described by the standard Gibbs ensemble but rather by a generalized
Gibbs ensemble (GGE) which has `chemical potentials' for each
conserved charge. There are a plethora of examples in string theory in
which integrable field theories appear, including many which have
gravity duals (see below). A question naturally appears whether we can
understand equilibration in these models via GGE, and if so, how to
interpret these in gravity.\footnote{In AdS/CFT and similar studies,
  thermal equilibrium has often been described in terms of 
  black holes, which normally possess only a {\it few} parameters. A GGE, with
  an infinite number of chemical potentials, presents a novelty {\it vis-a-vis}
this standard paradigm (see Section \ref{sec:discussions} for a detailed
discussion).}  
One of the purposes of this
paper is to initiate a study in this direction by studying Matrix
Quantum Mechanics (MQM) models, which appear in a variety of contexts
in string theory, as discussed below.

Besides the above issue, MQM offers simple toy models to study
dynamical phase transitions. Compared to some of the previously
studied statistical mechanics models, MQM has an advantage that it has
a simple large $N$ limit ($N$ is the rank of the matrix) which is
described by a semiclassical Fermi liquid. The dynamics of the
eigenvalue density of the matrix (whose gaps characterize the
thermodynamic phases) is inferred in this limit by the classical
motion of these droplets. In this paper, we study dynamics of a
quantum quench, both for finite $N$ (numerically) and in the
semiclassical large $N$ limit. 

The MQM (matrix quantum mechanics) we will study is the
following \footnote{\label{ftnt:singlet}We will restrict ourselves to
  excitations in the singlet sector of the model. This can be
  precisely achieved \cite{Wadia:1980cp} by coupling to a gauge field $A_0$: $\del_t U \to
  D_tU$, ~$D_t = \del_t - i A_0$ or by restricting to low energy
  dynamics  \cite{Gross:1990ub} which cuts off the non-singlets. For our purposes, we may consider
  \eq{fermion-single} and \eq{fermion-double} as the definition of the
  single-trace and double-trace models respectively.} 
\begin{align}
Z &= \int DU(t) \ e^{iN S}, ~ S = \int \kern-5pt dt \left[\frac{1}{2} 
\Tr \left(| \del_tU|^2 \right)  
-V(U) \right], 
\label{unitary-action}
\end{align} 
We will focus on two cases:
\begin{align}
&\hbox{(i) the single-trace model, given by~~} V(U) =\frac{a}{2}  
\left( \Tr U + \Tr U^\dagger \right), 
\label{single-trace}\\
&\hbox{and (ii) the double-trace model, given by~~}
V(U) =-\frac{\xi}{N} (\Tr U)(\Tr U^\dagger).
\label{double-trace}
\end{align}
Here $U(t)$ is a unitary matrix of rank $N$. The model
\eq{single-trace} was first introduced in \cite{Wadia:1980cp}
in the context of 2+1 dimensional gauge theory.
We will also discuss a
related hermitian matrix model \eq{hermitian} which shares similar
universal features with \eq{single-trace}. 
We will discuss, in Section \ref{sec:discussions}, string theory/gravity duals of these
models, in particular the connection between (\ref{single-trace}) and the non-critical 2D string theory, and (\ref{double-trace}) and the
near-horizon D2 brane geometry\footnote{For other, recent,
discussions of quantum
quench in AdS/CFT, see \cite{Das:2011nk} and references therein.}.

We list below a few highlights of our results\footnote{\label{ftnt-movie}
The movies for our time evolutions are available on\\
 \url{http://www2.yukawa.kyoto-u.ac.jp/~mtakeshi/MQM/index.html}
}:

(i) We find evidence of equilibration\footnote{For some other works on thermalization in 
matrix models, see Refs \cite{Asplund:2011qj, Riggins:2012qt, Asplund:2012tg}, which
consider BFSS and pp-wave matrix models.} in the time-evolution of certain observables ({\it e.g} moments
of the eigenvalue distribution) in both the MQM examples above.  In
case of the single-trace models, which is integrable, we find
excellent agreement of the equilibrium configuration with that
predicted by GGE (see Fig.~\ref{fig-rho1-gapless-gapped-single}). We
also give simple examples of observables which do {\it not} equilibrate.

(ii) We are able to give an interpretation of the equilibration in a
special case where an initial long wavelength perturbation dissipates
into progressively shorter and shorter wavelength perturbations, as in
the Richardson-Kolmogoroff cascade in turbulence. (See Fig.~\ref{fig:cascade}).

(iii) A surprising conclusion of our semiclassical analysis of the
quantum quench is that dynamical transitions from the gapless phase to
a gapped phase is possible, but not the other way around. In terms of
the semiclassical Fermi liquid description (see Section
\ref{sec:1-way}) the interpretation of this is that Fermi liquid
droplets cannot split.  In the gravity context, such as in a
Gregory-Laflamme transition between a black string and a black hole,
these observations have a potential relation to a change of topology of
the horizon. See Section \ref{sec:topology} for more details. 

The rest of the paper is organized as follows.  
  In Section
  \ref{sec:single-trace} we discuss the integrable single trace model
  and its dynamics, including equilibration and GGE, entropy
  production, energy cascades and dynamical phase
  transitions. In Section \ref{sec:double-trace} we discuss the double
  trace model, including its relation to gauge theory, dynamical phase
  transitions and equilibration/oscillations for various initial
  conditions.  In Section \ref{sec:discussions} we discuss implication
  of our results for the string theory/gravity duals of these models
  and possible generalization to a larger class of models.  In
  Appendix \ref{app:droplet} we discuss the fermion picture of the
  matrix models, and the semiclassical Fermi liquid description at
  large $N$.  In Appendix \ref{app:GGE} we discuss the notion of the
  generalized Gibbs ensemble in integrable models and its application
  to the single trace matrix quantum mechanics. In Appendix
  \ref{sec:QQ} we discuss the formalism of quantum quench dynamics in
  these models, and present explicit formulae for computing various
  relevant time-dependent quantities. In Appendix \ref{app:free} we
  discuss a common limiting case of both the single-trace and double-trace
  models in which the fermions are free and are without a potential; the
  density profile has a nontrivial nonlinear dynamics and shows, in this case, a power law decay
  to equilibrium. In Appendix \ref{app:details} we present some details
of our numerical analysis used to compute time evolution.

\section{\label{sec:single-trace}Single trace model}

The single trace model \eq{single-trace}, as explained in Appendix
\ref{app:droplet} and in the references cited there, can be
equivalently formulated in terms of $N$ noninteracting fermions moving
in a one dimensional spatial circle with an external potential $V(\t)= a \cos\t$.
The hamiltonian, in the second quantized form, is (see
\eq{fermion-single}) 
\begin{align}
H =&
\int_{-\pi}^\pi \kern-5pt d\theta\ \psi^\dagger(\theta,t)
h \psi(\theta,t), \qquad h=- \frac{1}{2N^2}
  \del_\t^2 + V(\t), \qquad
V(\t)= a \cos\theta.
\label{hamiltonian-single}
\end{align}
The fermion density
\begin{align}
&\rho(\t,t)=  \frac1N \psi^\dagger (\t,t) \psi(\t,t), \qquad
 \int_{-\pi}^\pi \kern-5pt d\t\ \rho(\t,t)= 1, \nonumber
\end{align}
 is the same
as the eigenvalue density of $U(t)$. 
Depending on the parameter
  $a$, there are two different static phases (ground states) in the
  system at large-$N$: (i) gapless for $a< a_c = \pi^2/64$, and (ii) gapped for
  $a>a_c$, where the `gap' refers to that in the equilibrium value of
  $\rho(\t)$ (see Fig.~\ref{fig:gap-or-not} for various shapes of $\rho(\t)$). 
  The quantum critical transition at $a_c$ is a 
  third-order transition  \cite{Wadia:1980cp} \footnote{This is similar to
the Gross-Witten-Wadia transition  \cite{Gross:1980he, Wadia:2012fr}
of zero-dimensional single-trace unitary matrix model.}
  (see below \eq{scaling}).

In what follows, we will analyze various types of dynamics of the
above system: 
\\ (i) We study time evolution of some arbitrarily
chosen initial state (corresponding to a suitable density profile)
under the hamiltonian \eq{hamiltonian-single}.  
\\ (ii) Quantum Quench
Dynamics (QQD) (see Section \ref{sec:QQ}) for details): we start, in
the far past, with the ground state of the Hamiltonian
\eq{hamiltonian-single} with an initial parameter $a=a_i$; we then
change $a$ suddenly to $a_f$ at $t=0$, and study the $t\ge 0$
evolution of this initial state and various expectation values under
the hamiltonian \eq{hamiltonian-single} with parameter $a=a_f$. We can
regard this as a special case of (i) where the initial state is
provided by the ground state of the system in the far past.

In this section, we just show the results of these dynamics, and
the details of the calculations are summarised in appendixes.

\subsection{\label{sec:GGE-text}Integrability and equilibration}

Since the hamiltonian \eq{hamiltonian-single} describes free fermions,
it is clearly integrable. 
Indeed we can find the conserved quantities as follows
(see Appendix \ref{app:GGE} for details).
Let us expand the fermion field as 
\begin{align}
\psi(\theta)=\sum_{m=0}^{\infty} c_m \varphi_m (\theta), \quad
 h \varphi_m (\theta) = \epsilon_m \varphi_m,
\end{align}
where $\varphi_m$ is the $m$-th eigenfunction of the single particle hamiltonian $h$ \eq{hamiltonian-single} with the eigenvalue $\epsilon_m$ and $c_m$ is the corresponding annihilation operator. The fermion occupation number operators
\begin{align}
N_m = c^\dagger_m c_m, \quad (m=0,1,\cdots ) ,
\label{fermion-number}
\end{align}
are clearly all conserved, and all independent 
in the $N\to\infty$ limit.


A first guess about a free system such as this would be that no
observable can possibly equilibrate since there presumably cannot be
any dissipation. Indeed, such a guess would seem to have a lot of
merit. Firstly, there are infinite number of charges, such as the
$N_m$, which, given any initial state however complicated, clearly
stay constant as the system evolves. Secondly, it is easy to construct
a whole series of observables which do not remain constant but keeps
oscillating for ever; as an example consider the operator
\begin{align}
\hat O_{m,n}(t) = c^\dagger_{m} c_n \exp[-i(\ep_{m} -
  \ep_{n})t/\hbar], \quad m\ne n.
\label{non-equil}  
\end{align}
Clearly, given {\it any} state $|
\Psi(t) \ran$, the expectation value of the above operator evolves as
(using the Heisenberg picture)
\begin{align}
\vev{\hat O_{m,n}(t)} & \equiv \lan \Psi(0) | \hat O_{m,n}(t) |
\Psi(0) \ran = V_{m,n} \exp[-i(\ep_{m}- \ep_n)t/\hbar],\quad
\label{oscillating}
V_{m,n}  \equiv \lan \Psi(0) |c^\dagger_{m} c_{n}  | \Psi(0) \ran.
\end{align}
Thus, the expectation value keeps oscillating {\it ad infintum}
(unless $V_{m,n} = 0$ in the given initial state, in which case the
value remains zero).
It is obvious how to
generalize such examples, e.g. by combining many creation-annihilation
pairs.

\paragraph{Behaviour of $\rho(\t,t)$:}

In the light of the above discussion, we find surprises when we
consider the time evolution of the density $\rho(\t,t)$ or its various
`moments', defined by 
\begin{align}
\vev{\rho_n(t)} = \int  d\theta \vev{\rho(\theta,t)}
\cos(n\theta).
\label{moments}
\end{align} 
In \eq{rho-t} a finite-$N$ expression for $\vev{\rho( \theta,t)}$ is
presented for a quantum quench (defined briefly below
\eq{hamiltonian-single} and discussed in detail in Section
\ref{sec:QQ}). Using this expression, it is easy to compute the
time-dependence of various moments $\rho_n(t)$. 
 In Fig.~\ref{fig-rho1-gapless-gapped-single} {\bf (a)} we have plotted $\rho_1(t)$ for $N=120$. 
  In the same figure we have also plotted the $N\to \infty$
limit of $\rho_1(t)$, computed using \eq{moments} and the expression
\eq{rho-t-coll}, the latter being derived using the large $N$
semiclassical limit of the fermion theory \eq{hamiltonian-single}.
From these plots, the evidence of equilibration is quite clear; while
for finite $N$, there appears to be relaxation followed by a
recurrence (with the recurrence time increasing with
$N$ as shown in Fig.~\ref{fig-rho1-gapless-gapped-single} {\bf (b)}), 
at large $N$, $\rho_1(t)$ appears to relax to a certain value.

\begin{figure}[H]
\begin{center}
\begin{tabular}{cc}
\begin{minipage}{0.5\hsize}
\begin{center}
\includegraphics[scale=.4]{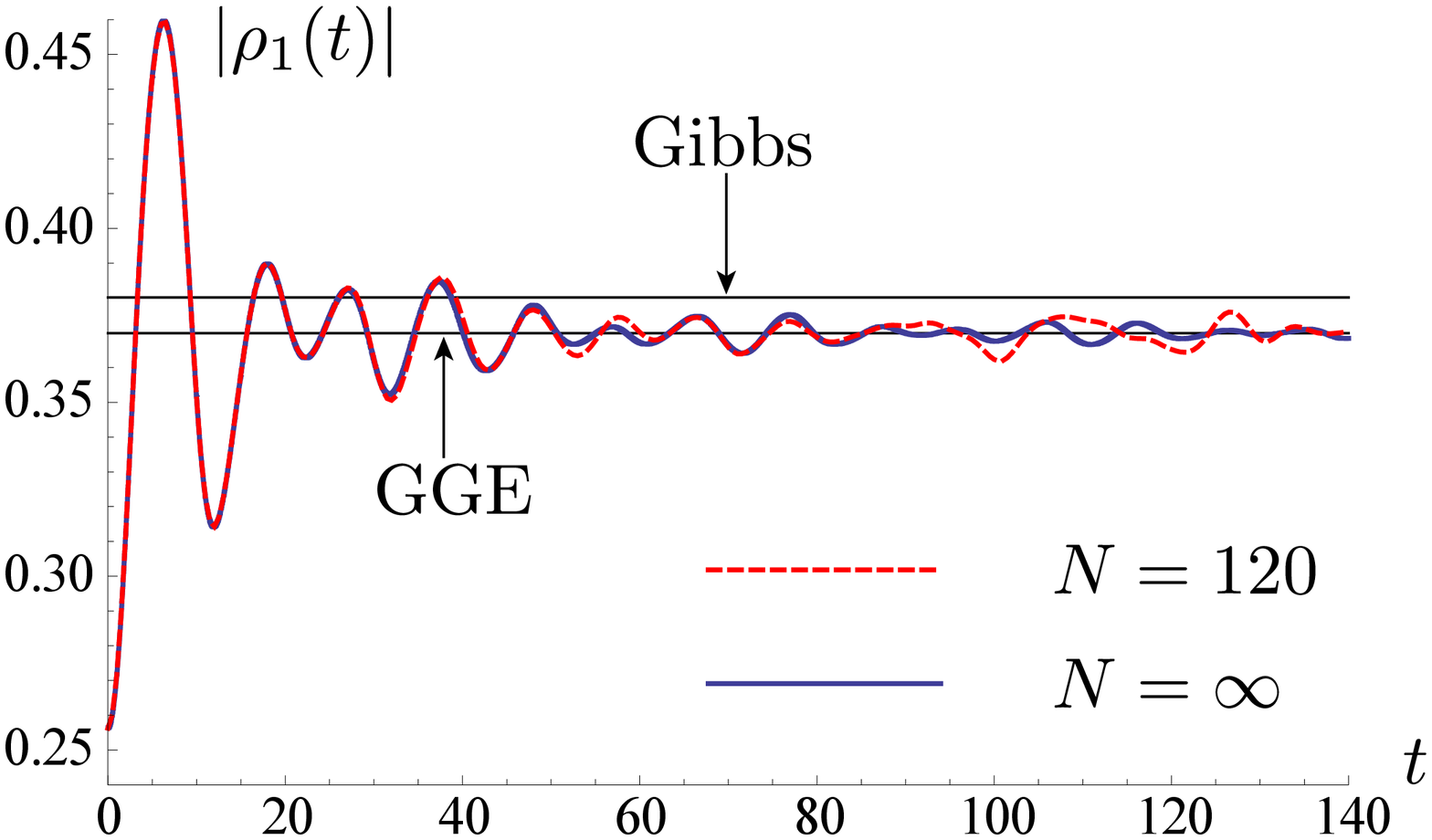}
\begin{center}
{\bf (a)} $\rho_1(t)$ and GGE vs. Gibbs ensemble
\end{center}
\end{center}
\end{minipage} 
\begin{minipage}{0.5\hsize}
\begin{center}
\includegraphics[scale=.4]{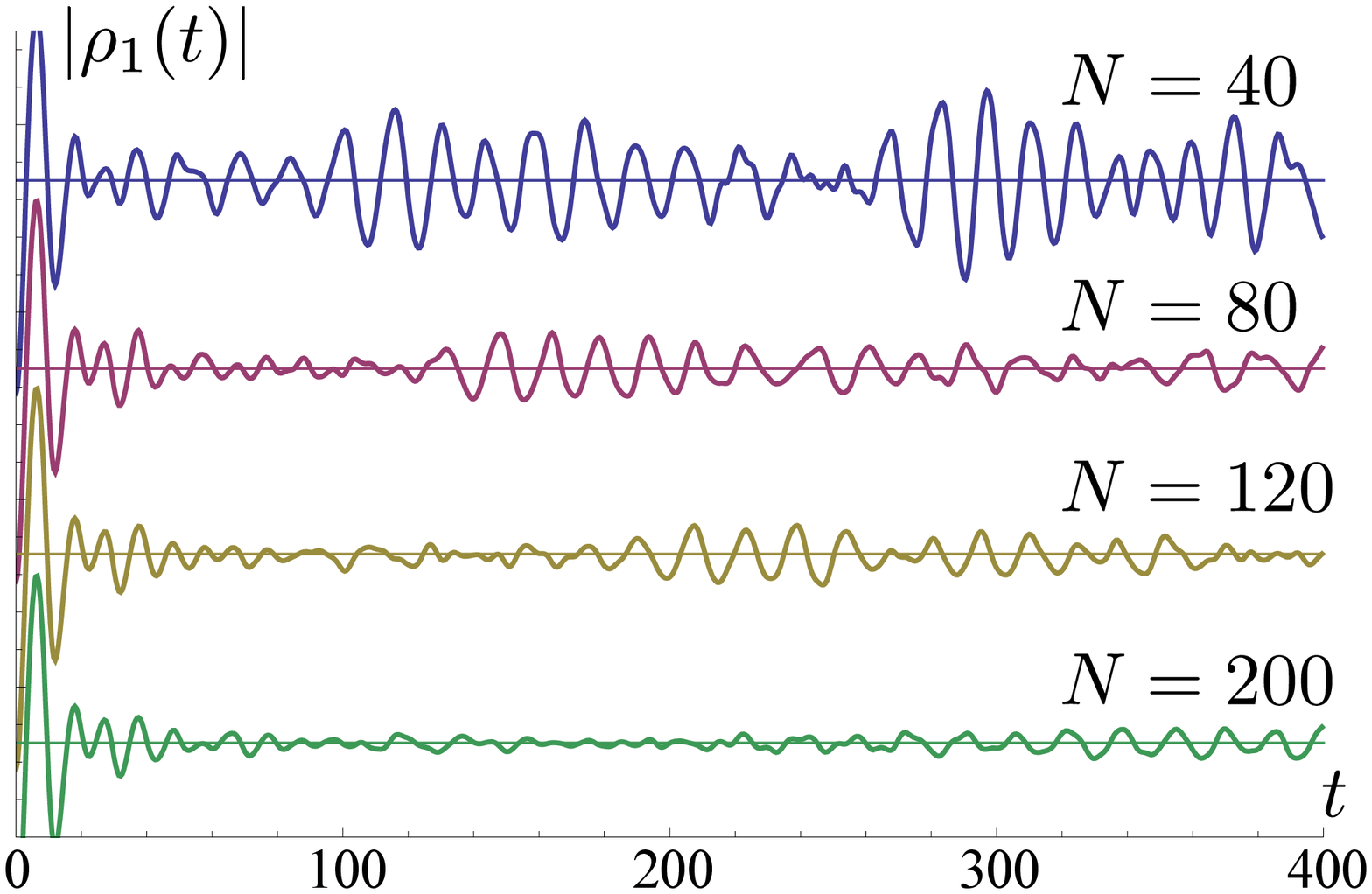}
\begin{center}
{\bf (b)} Poincar\'{e} recurrence
\end{center}
\end{center}
\end{minipage} 
\end{tabular} 
\caption{\footnotesize
 The time evolution of the first moment
  $|\vev{\rho_1(t)}|$ (defined in \eq{moments}) for a quantum quench is shown.
  We take $a_i = 0.8 a_c$ initially and change it to $a_f =1.2 a_c$ instantaneously at $t=0$.
  {\bf (a)} The (red) dashed line shows the result for $N=120$, which appears to show  Poincare cycles.
The Poincare cycles seem to grow linearly with $N$ as shown in {\bf (b)}.
The (blue) solid line in {\bf (a)} is for $N=\infty$ (computed using the droplet formalism, see Section \ref{app:droplet} and \ref{app:details}); it shows equilibration and absence of Poincare cycles. 
The mean value of $|\vev{\rho_1(t)}|$  for finite $N$ and its asymptotic value
 for $N=\infty$ are seen to have excellent agreement with each other and, in turn,  with 
  $|\vev{\rho_1}_{\rm GGE}|$ as computed in GGE, (see Section \ref{app:GGE} for details of the computation), whereas these values all
  differ significantly from the ensemble average in the 
  standard Gibbs ensemble. We have verified that the $N$ dependence of the 
$\vev{\rho_1}_{\rm GGE}$  is small.}
\label{fig-rho1-gapless-gapped-single}
\end{center}
\end{figure}

\paragraph{Understanding of the equilibration in terms of GGE:}
Since in view of the discussion around Eqn. \eq{oscillating}, the
above equilibration is surprising, it is useful to look for similar
phenomena in other examples. Fortunately, in recent years several
examples of selective equilibration have been found in integrable
systems (see \cite{Polkovnikov:2010yn} for a review), in which a
certain class of observables has been shown to equilibrate in such
integrable systems (which include the hard core bosonic lattice and
the transverse field Ising model), and the equilibrium configuration
is characterized not in terms of the standard thermal (Gibbs)
ensemble, but by a Generalized Gibbs Ensemble (GGE) which keeps track
of the infinite number of conserved quantities by means of an infinite
number of chemical potentials.\footnote{The infinite number of chemical
potentials can sometimes be thought of as a separate `temperature'
for every mode, see, e.g. \cite{Calabrese:2011GGE}.}
In our present model, the GGE is defined by the  density matrix
\begin{align*}
\varrho_{\rm GGE}= \frac1{Z_{\rm GGE}}\exp\left(-\sum_m \mu_m N_m \right) ,
\end{align*}
where $\mu_m$ is the chemical potential corresponding to the
conserved fermion number $N_m$ (\ref{fermion-number})
  (see sections \ref{app:GGE} and
\ref{sec:GGE-QQ} for  details). 
Using this density matrix, we are able to compute the postulated
 equilibrium
value of the density $\vev{\rho(\t)}_{\rm GGE}$ = $\Tr \left(\varrho_{\rm GGE} \rho(\t)
\right)$ and hence, through \eq{moments}, that of the various moments $\vev{\rho_n}_{\rm GGE}$. 
 In Fig.~\ref{fig-rho1-gapless-gapped-single} we have shown the value of
$\vev{\rho_1}_{\rm GGE}$; we find that the large $t$ asymptotic value of $\vev{\rho_1(t)}$
computed at large $N$ fits very well with the GGE-value. For
comparison, we have also displayed in the same figure the  value of
$\rho_1$ computed according to the Gibbs ensemble, which differs
significantly from both the asymptotic  value of $\vev{\rho_1(t)}$
as well as $\vev{\rho_1}_{\rm GGE}$. We have verified these statements
for higher moments $\rho_2, \rho_3$ etc. as well.  We find,
therefore, that the moments $\rho_n(t)$ belong to the class of
observables in our system which equilibrate\footnote{The present
  example provides evidence for relaxation for a quantum quench
  dynamics. However, we will find later the same phenomenon with other
  initial conditions; in particular see Section \ref{app:free} where
  we provide an analytic proof of such relaxation in a special case.},
and the 
equilibrium configuration can be quantitatively understood in terms a
GGE. 

So far we have seen two classes of observables.
One, like \eq{non-equil}, never equilibrates and another, like  $\rho_n$
above, equilibrate to values described by GGE.
It is, therefore, a subtle issue whether the whole system should be regarded as equilibrated or not.
Indeed the criterion for equilibration in quantum systems has a long history 
(see, e.g. \cite{Polkovnikov:2010yn} for a recent history).
One way to define it is considering only macroscopic quantities\footnote{Another way is to define it in terms of long time averages of macroscopic quantities,
as in the discussion of ergodic motion.}.

In our model, we should regard observables like (\ref{non-equil}) as microscopic, since they connect a given energy eigenstate to a few (only one 
in the precise example of \eq{non-equil}) eigenstate(s)
 , and should regard $\rho_n$ as macroscopic, since they connect
a given eigenstate to an infinite number of other eigenstates (these
observables are also moments of the fermion density which is clearly
a macroscopic, collective, field). 
It appears, therefore, that the following equation is 
satisfied for macroscopic observables $O$
\begin{align}
\Tr(\varrho_{\rm true}~ O(t)) \to \Tr(\varrho_{\rm GGE} ~O(t)), \quad \hbox{as}~ t\to
\infty,
\label{GGE-hypo-new}
\end{align} 
where $\varrho_{\rm true} = |\Psi(0) \ran \lan \Psi(0) | $ is the density
matrix corresponding to the initial state at $t=0$. 
Thus, according to the above criterion, we conclude that our model equilibrates to GGE\footnote{Usually a criterion like eq.(\ref{GGE-hypo-new}) is used for the standard Gibbs ensemble. Here we are proposing a generalization
of that to the generalized Gibbs ensemble.}.

\subsection{\label{sec:cascade}Energy Cascade: an interpretation
of the dissipation in the integrable model}

To get more insight into the phenomenon of equilibration, let us
consider the special case of $V=0$ (i.e. $a=0$) in
\eq{hamiltonian-single}. In stead of a quantum quench type initial
configuration, let us in this case start with an 
initial state $| \Psi_0 \ran$ which
corresponds to a sinusoidal semiclassical configuration (see
\eq{collective-single} for notation)\footnote{This kind of
initial condition is chosen here only for simplicity; we expect
the same qualitative behaviour for the quantum quench initial conditions 
as well.}
\begin{align}
\rho(\t,t=0)=  \frac{1}{2\pi}\left(1 + 2 b \cos(n \theta) \right),
~ {\cal P}(\t,t=0)=0.
\label{ini-density}
\end{align}
The initial condition implies that only $\rho_n$ is excited at $t=0$ with an amplitude $b$. 
By using the technique detailed in Appendix \ref{app:free}, we can
compute the time evolution of the system (see Figure
\ref{fig-thermalization}), and compute expressions for the moments
$\rho_n(t)$ analytically. For instance, the late time behaviour of the
first two moments (for the choice $n=1$ in \eq{ini-density}) is given
by
\begin{align} 
\rho_1(t) \to \frac{2}{t^{3/2}} \sqrt{\frac{2 }{ \pi b} } \cos \left(  
bt-\frac{3\pi}{4} \right)   \cos\left( \frac{t}{4\pi}\right) , \quad (t 
\to \infty),
\label{late-rho1}
\end{align} 
and 
\begin{align} 
\rho_2(t) \to  \frac{1}{t^{3/2}} \frac{1}{ \sqrt{\pi b }} \cos\left( 2bt-
\frac{\pi}{4} \right)  \sin\left( t\right) , \quad (t \to \infty).
\label{late-rho2}
\end{align}
Indeed it turns out that ALL the moments die out as $t^{-3/2}$ (see
Fig.~\ref{fig:cascade}).\footnote{The power-law relaxation may be a
  special feature of the simple $a=0$ model.  In the more general
  case, such as in Fig.~\ref{fig-rho1-gapless-gapped-single}, the
  relaxation appears to have a characteristic time scale.}

\begin{figure}
\begin{center}
\includegraphics[scale=0.4]{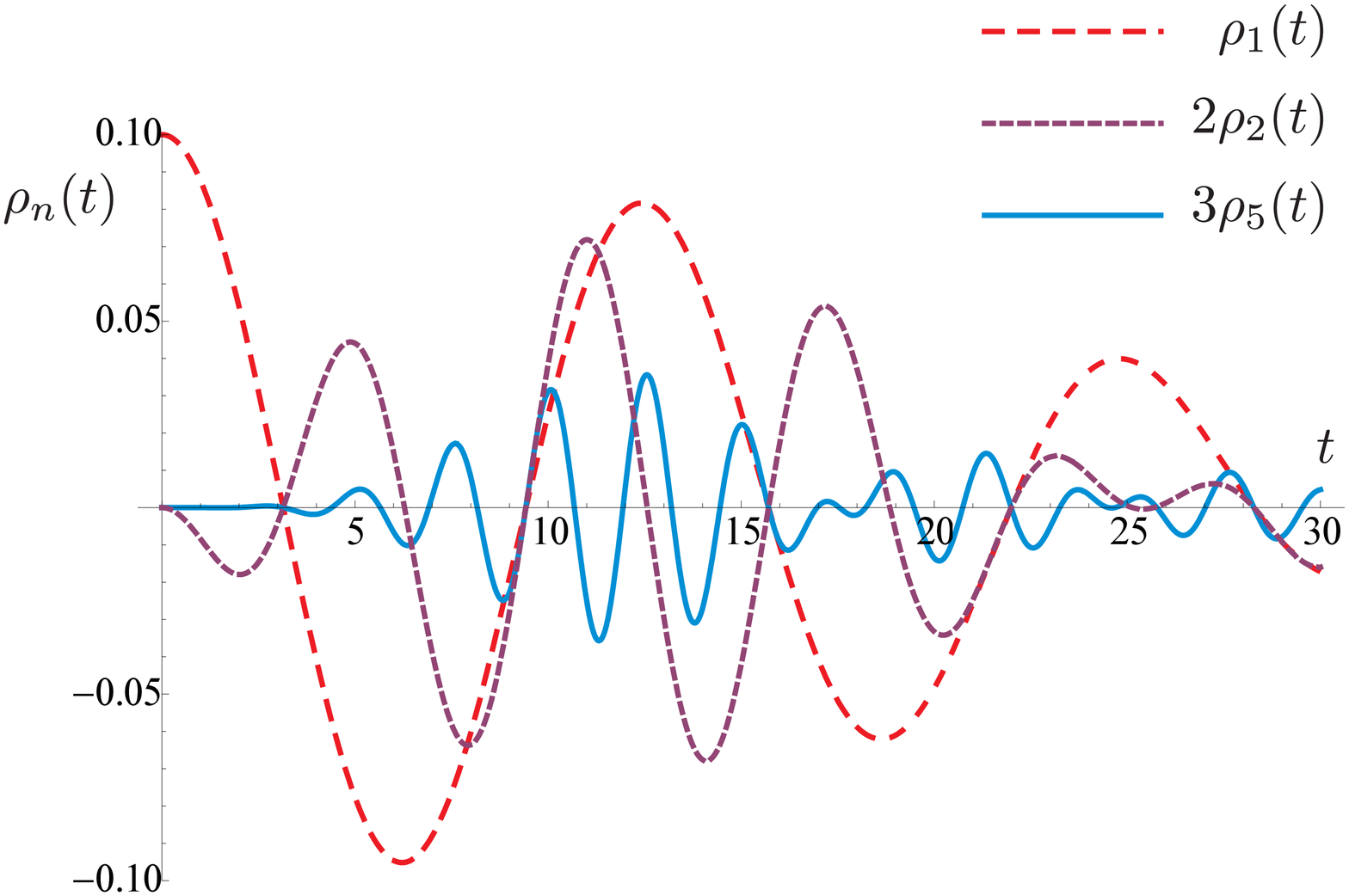}
\caption{\footnotesize  Time evolution of moments of the eigenvalue density for the free model (see Appendix \ref{app:free} for
details). Note the time of appearance of the first
  peaks.  Lower frequency modes get excited and die out first. 
Higher frequency modes get excited and die out later. There is a flow of energy
from longer wavelength modes to shorter wavelength modes, as in
the Richardson-Kolmogoroff cascade in turbulent fluids.}
\label{fig:cascade}
\end{center}
\end{figure}

It would appear surprising at first sight that ALL the moments die away!, leaving us apparently with the question `where does the energy go?'  Of
course, since we have a conservative model, even integrable to boot,
energy cannot really `go' anywhere. It turns out, just as studies
in turbulence and various other areas have taught us, that there is a flow
of energy from one set of `modes' to another set of `modes'. To be
precise, as the longer wavelength modes die, shorter wavelength modes
start getting excited, and as they too die away, even shorter
wavelengths get excited, etc. This phenomenon is similar to the
Richardson-Kolmogoroff type cascade in turbulence, and provides a
mechanism of dissipation in our integrable system.

\subsection{\label{sec:entropy}Entropy} 

As we remarked above, our model equilibrates, in the sense of
eq.(\ref{GGE-hypo-new}).  Since the initial density matrix
$\varrho_{\rm true}$ in that equation describes a pure state, with
zero von Neumann entropy, while the final density matrix on the right
hand side describes a mixed state, with non-zero von Neumann entropy
(see \eq{GGE-entropy}), we can say that a non-zero entropy has been
generated during the time evolution.

We should, of course, clarify here that \eq{GGE-hypo-new} 
cannot be interpreted to mean $\varrho_{\rm true}(t)$ $\equiv
e^{-iHt} \varrho_{\rm true} e^{iHt}$~~~$  \stackrel{t\to\infty}{\to} 
\varrho_{\rm GGE}$~! The
correct interpretation of \eq{GGE-hypo-new} is that the macroscopic
observables $O(t)$ are suitably coarse-grained to smooth out the
difference between $\varrho_{true}$ and $\varrho_{\rm GGE}$ after a sufficiently
long time. It is in these sense that we can regard that the entropy $S_{\rm GGE}$ has been produced.

\paragraph{Amount of entropy production:} 
The von Neumann entropy of the final state,
$S_{\rm GGE}$ = $-\Tr~ \varrho_{\rm GGE} \ln 
\varrho_{\rm GGE}$, is computed in \eq{GGE-entropy}. 
Using this equation 
and \eq{N-m-sudden} we can compute the entropy $S_{\rm GGE}$
characterizing the GGE for a quantum quench. In Fig.~\ref{fig:entropy}, 
we show the result of this computation as a function of $N$. 
From the plots we
can see that $S_{\rm GGE} \propto N$.\footnote{We have also verified that
in units of \eq{hamiltonian-single}, the conserved energy of the system 
is $E = O(N)$ and the temperature is $T= O(1)$.}  
 This could be roughly expected from the
fact that there are $N$ non-interacting particles in the system. We discuss
below an analytic calculation of the entropy. 
In Section
\ref{sec:discussions} we discuss this $O(N)$ entropy in terms of various
string theory duals. 

\begin{figure}
\begin{center}
\begin{tabular}{cc}
\begin{minipage}{0.5\hsize}
\begin{center}
\includegraphics[scale=.4]{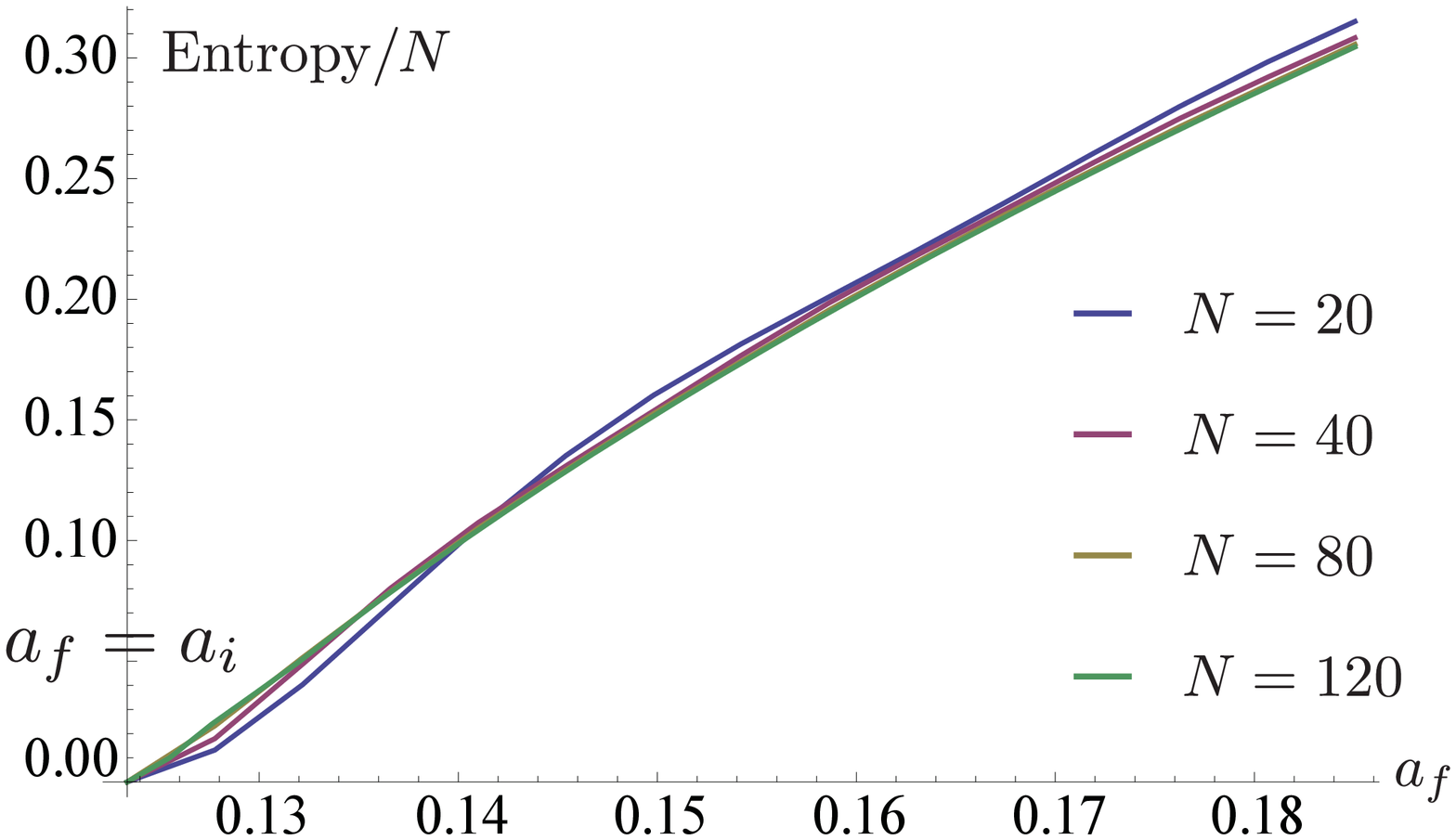}
\begin{center}
{\bf (a)} $a_i<a_c$
\end{center}
\end{center}
\end{minipage} 
\begin{minipage}{0.5\hsize}
\begin{center}
\includegraphics[scale=.4]{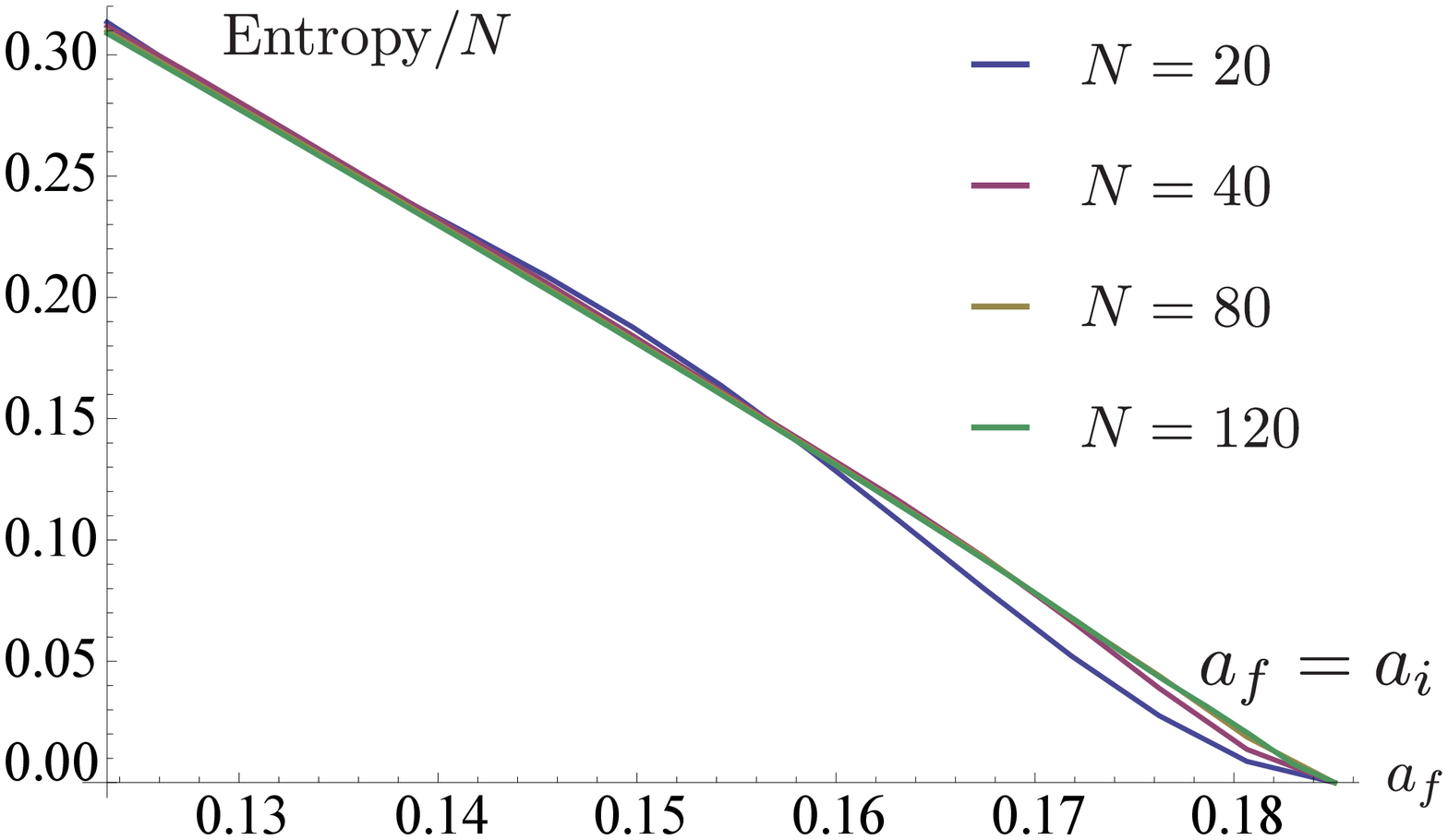}
\begin{center}
{\bf (b)} $a_i>a_c$
\end{center}
\end{center}
\end{minipage} 
\end{tabular} 
\caption{\footnotesize
 $S_{\rm GGE}/N$ vs $a_f$  for quantum quenches. These results indicate that $S_{\rm GGE}$ scales with $N$.
 We can also see that they are proportional to $|a_i-a_f|$. }
\label{fig:entropy}
\end{center}
\end{figure}

\paragraph{\label{para:analytic-S}An analytic calculation of etropy} 
We will now present an analytic, large $N$, computation of the entropy
$S_{\rm GGE}$ in the $V=0$ case argued in section \ref{sec:cascade}.
As shown in Appendix \ref{app:free}, we can evaluate the entropy as
\begin{align}
S=  2.36\ b\ N
\end{align}
 Note that $S$ is exactly linear in $N$ (as well as in the amplitude $b$
of the initial perturbation \eq{ini-density}).

\subsection{\label{sec:dyn-ph-tr}Dynamical phase transitions}

We mentioned above that if we tune the parameter $a$ in
\eq{hamiltonian-single}, say from zero upwards, we encounter a third
order quantum phase transition at $a= a_c = \pi^2/64$ from a
gapless phase to a gapped phase. What happens if we make this change
of $a$ {\it dynamically}, say from $a_i<a_c$ to $a_f> a_c$? If we prepared
the system in the ground state initially (corresponding to $a=a_i$),
does it eventually relax to the final ground state (corresponding to
$a=a_f$)?  Do we observe dynamical appearance of a gap at a certain
time? What role does the critical point $a_c$ play, if any? What
happens if we make the change from $a_i>a_c$ to $a_f<a_c$?

The general setup for addressing these questions is to consider a
dynamical variation with a tunable `ramp speed', e.g. in the
Kibble-Zurek transition (see, e.g., the review
\cite{Polkovnikov:2010yn}). However, in the present work, we will
consider two special cases, namely (a) the quantum quench dynamics\footnote{
In this section, we use GGE to evaluate the asymptotic values of the observables in the QQD, although,  at finite $N$, the actual values keep oscillating slightly around the GGE predictions as shown in Fig.~\ref{fig-rho1-gapless-gapped-single} {\bf (b)}. } discussed
in the previous section, in which the change from $a_i$ to $a_f$
happens instantaneously, and (b) the adiabatic case, in which
the change happens at a rate slower than the scale set by
energy level spacings\footnote{Since the energy level spacings become zero at large-$N$, the adiabatic approximation works only if we tune the change of $a$ slower than $O(N)$.}.  We will first address the
question of appearance/disappearance of a gap in the final state in
case we cross the critical value $a_c$ from an initially
gapless/gapped phase.

\subsubsection{\label{sec:1-way}One-way phase transitions} 

Fig.~\ref{fig:gap-or-not} shows, for specific values of $a_i, a_f$,
the large $t$ asymptotic density $\rho(\t,t)$ for quantum quench
dynamics (QQD), calculated from \eq{rho-t}, as well as the same
quantity for adiabatic dynamics (where the time-dependent wavefunction
is approximated by the instantaneous ground state), as calculated in
\eq{adiabatic-single}. We find that there is a rather remarkable
`time-asymmetry' in the QQD, {\it viz. that while a gapped $\to$
  gapless transition is possible, a gapless $\to$ gapped transition is
  impossible}. In other words, in the context of QQD, while a gap can
close dynamically, it cannot open dynamically.\footnote{Of course,
  this is not a {\it real} time-asymmetry in the sense that, for
  $a_i<a_c<a_f$ if we start from the rather complicated, gapless,
  final pure state, and reverse the direction of time, we will indeed
  get back the initial gapped pure state. However, in the context of
  QQD, we always start from a ground state, in either direction,
  leading to the apparent time-asymmetry.}

\begin{figure}
\begin{center}

\begin{tabular}{cc}
\begin{minipage}{0.5\hsize}
\begin{center}
\includegraphics[scale=.4]{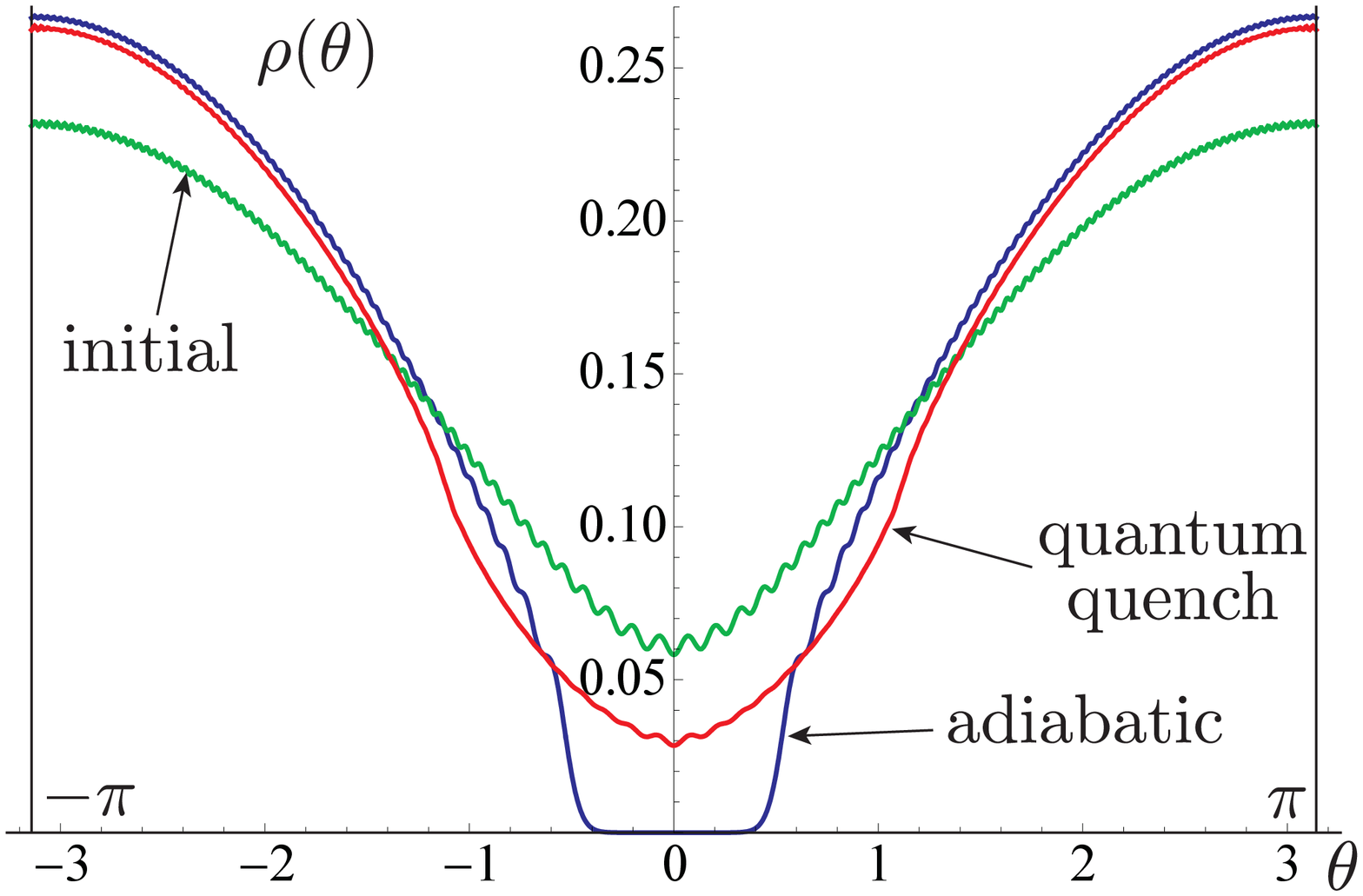}
\begin{center}
{\bf (a)} $a_i<a_c < a_f$
\end{center}
\end{center}
\end{minipage} 
\begin{minipage}{0.5\hsize}
\begin{center}
\includegraphics[scale=.4]{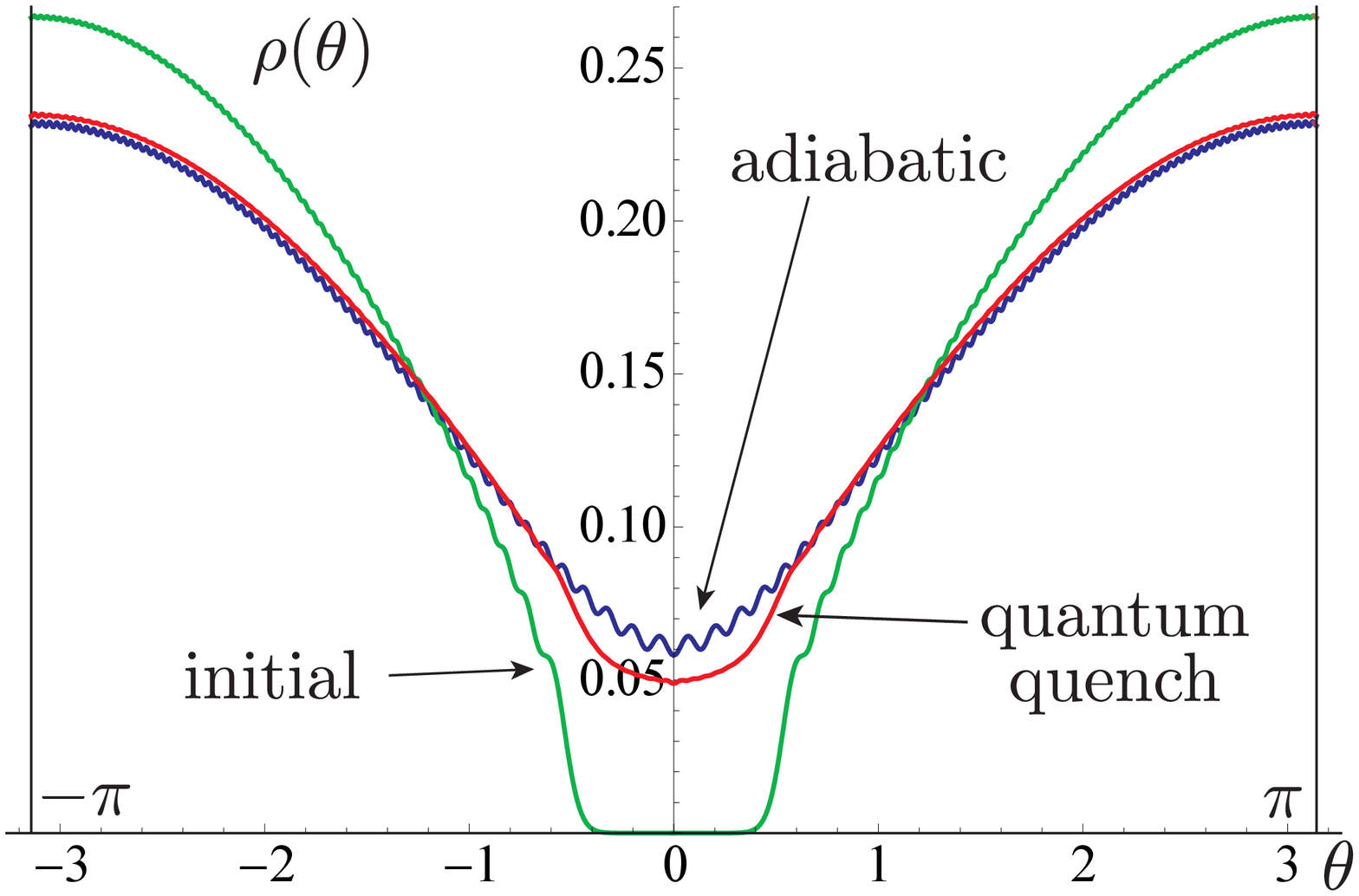}
\begin{center}
{\bf (b)} $a_i>a_c>a_f$
\end{center}
\end{center}
\end{minipage} 
\end{tabular} 
\caption{\footnotesize A plot of density $\vev{\rho(\t)}$ {\it vs}
  $\t$ at $N=120$. In the left panel, $a_f>a_c>a_i$; the green curve shows the
  initial gapless density as a function of $\t$,
   the blue curve shows
  the gapped density corresponding to the final ground state
  for $a=a_f$, and the red curve shows the large $t$
  asymptotic value of $\rho(\t,t)$ in quantum quench dynamics. 
 Since no gap appears, no dynamical phase transition occurs in the quench case. 
  In the
  right panel, $a_f < a_c < a_i$, the initial density is gapped,
  whereas both the final configuration in the actual dynamics as well
  as the final ground state show a gapless phase, showing the presence
  of a dynamical phase transition in this case.}
\label{fig:gap-or-not}
\end{center}
\end{figure} 

In Fig.~\ref{fig:no-gap} we present more details about the gap at
varying values of the $a$-parameter. As seen in
Fig.~\ref{fig:gap-or-not}, for $a>a_c$ the gap in the ground state
density $\rho(\t)$ opens around $\t=0$ (because the potential
$a\cos\t$ draws the fermions from $\t=0$ towards $\t=\pi$, as
explained below \eq{scaling}). Thus, the value of the density at
$\t=0$
can be regarded as an order parameter which is non-zero in the gapless
phase and zero in the gapped phase\footnote{These statements refer to
  the large $N$ limit. For finite $N$, or in a double scaled limit, a
  non-zero fermion density exists in the `gapped region'.}. In
Fig.~\ref{fig:no-gap}, we present this order parameter for various
values of $a_f$ for QQD after relaxation and compare it with the
ground state. The one-way nature of transition is quite clear in this
plot as well. Note also the progressive departure from adiabaticity as
$a_f$ is varied, starting from $a_i$, across $a_c$ to the other phase
(this is more prominent in the left panel where we study the
appearance of a gap). We will have more to say shortly about the
departure from adiabaticity. 

\begin{figure}[H]
\begin{center}
\begin{tabular}{cc}
\begin{minipage}{0.5\hsize}
\begin{center}
\includegraphics[scale=.4]{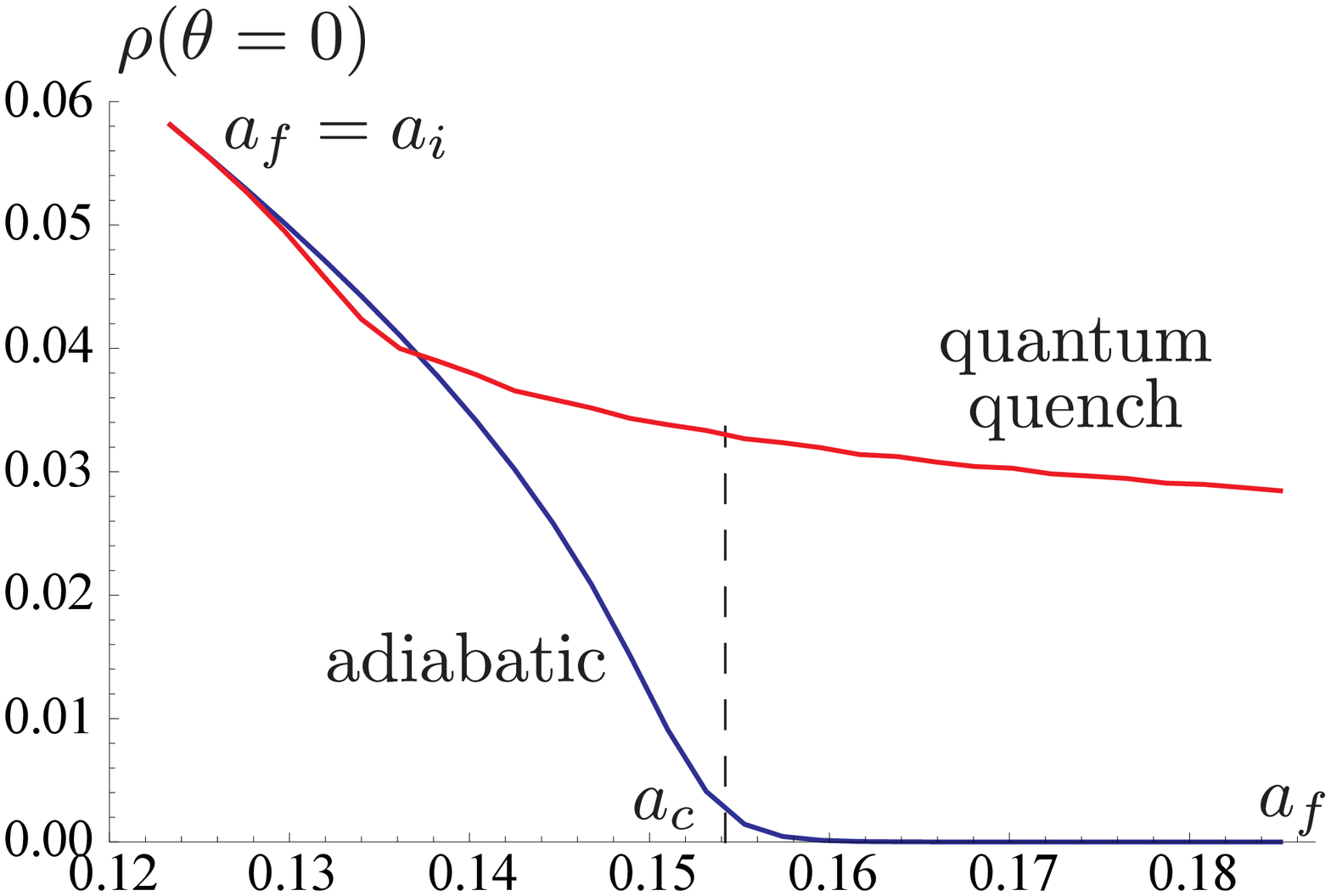}
\begin{center}
{\bf (a)} $a_i<a_c $
\end{center}
\end{center}
\end{minipage} 
\begin{minipage}{0.5\hsize}
\begin{center}
\includegraphics[scale=.4]{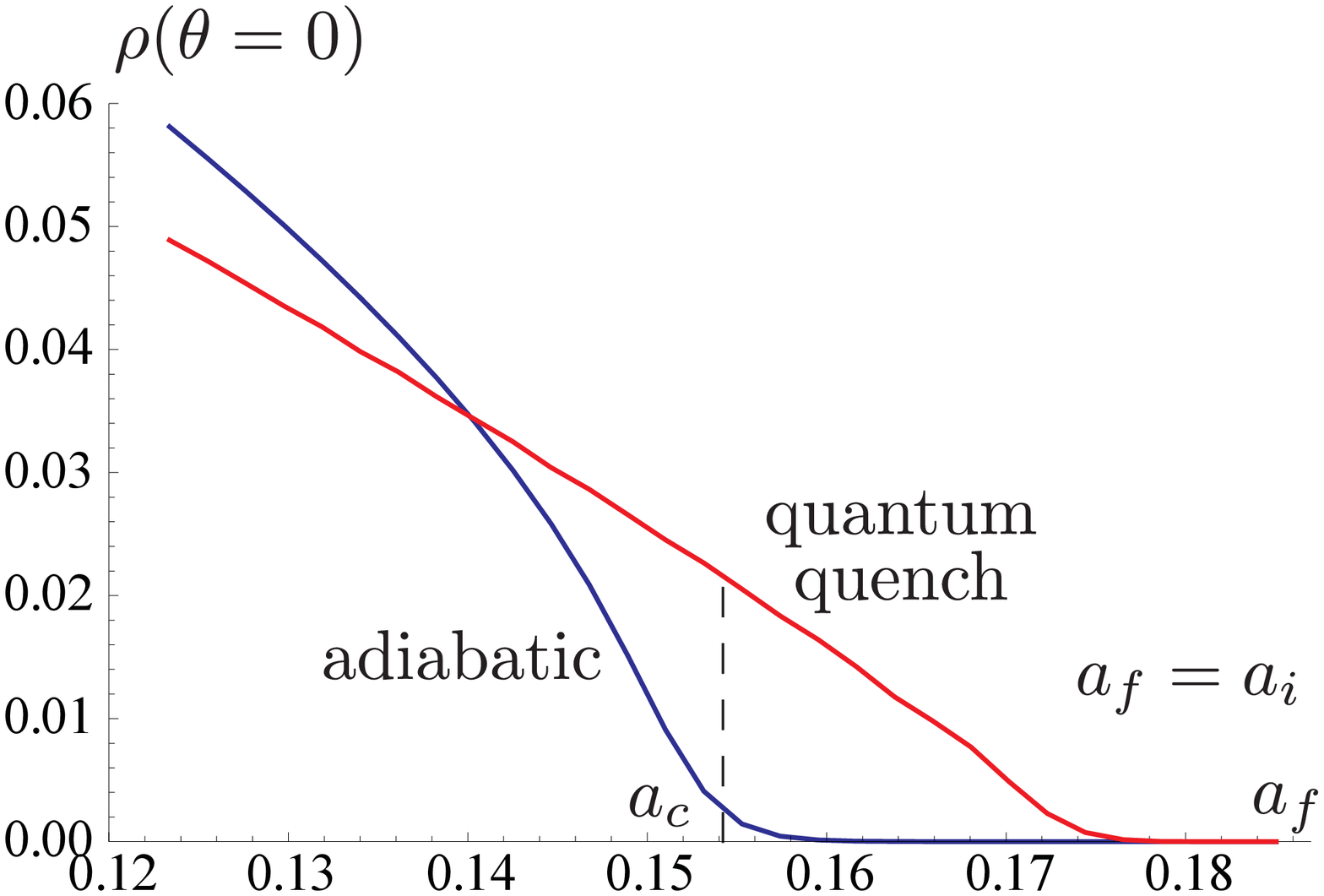}
\begin{center}
{\bf (b)} $a_i>a_c$
\end{center}
\end{center}
\end{minipage} 
\end{tabular} 
\caption{\footnotesize A plot of $\rho(\theta=0)$ for various values of
  $a_f$ at $N=120$. In the left panel, we begin in the gapless phase with
  $a_f=a_i < a_c$, and consider a continuous sequence of QQD
  evolutions in which $a_f$ is progressively increased across
  criticality to $a_f>a_c$ : the `adiabatic' curve, calculated
  according to \eq{adiabatic-single}, shows transition to a gapped
  phase (with $\rho(\theta=0) \to 0$), whereas the actual dynamical evolution
  under quantum quench dynamics (QQD) shows that no gap appears
  dynamically ($\rho(\theta=0)$ remains non-zero). In the right panel, both
  the adiabatic and QQD curves show disappearance of the initial gap,
  confirming the presence of a dynamical {\it gapped $\to$ gapless}
  transition.
  Note that the gap does not disappear at $a_c$ even in the ground state (adiabatic case). This is due to a finite $N$ effect.
  }
\label{fig:no-gap}
\end{center}
\end{figure} 

\subsubsection{A topological interpretation of the one-way transition 
from Fermi liquid drops}
\label{sec:1-way-expl}

The apparent irreversibility in dynamical phase transitions, found
above, turns out to have a beautiful semiclassical interpretation from
the semiclassical Fermi liquid picture. For simplicity, let us
consider the large $N$ limit of the single-trace hermitian model
\eq{hermitian}, which corresponds to replacing the potential $a\cos\t$
in \eq{hamiltonian-single} by $-a\t^2/2$ (defined together with a hard
cut-off $\t = \pm \t_m$). Since in these models $\hbar=1/N$, the large
$N$ limit, given by \eq{scaling}, is given by the semiclassical
picture of a Fermi liquid in phase space, which is described by a
phase space density $u(\t,p,t)$ which is either 0 or 1, and the filled
regions $u=1$ are described as `droplets'. There is a simple
description of quantum quench dynamics in this picture, as already
alluded to in Section \ref{sec:GGE-text} and described in detail in
Section \ref{app:droplet}.

As we describe in Fig.~\ref{fig:dyn-phase-tr}, the time evolution from
the gapless phase towards the gapped phase involves starting with a
single connected Fermi droplet in the Fermi sea.  The evolution of the
droplet follows the equation \eq{u-hermitian}. It is not difficult to
see that the hyperbolic motion of the fermions (in which fermions
belonging to different hyperbolas move at different speeds) leads to
squeezing and stretching of the droplet, although it can never split.
Since $\rho(\t,t) = \int \frac{dp}{2\pi} u(\t,p,t)$, which can be
interpreted as a kind of projection on to the $\t$-axis, the
impossibility of droplet splitting in phase space translates to the
impossibility of a gap-opening transition in $\t$-space. In other
words, a continuous $u(\t,p,t)$ means that there will always be some
fermions in the phase space along the $p$-axis, hence $\rho(\t=0)$
will remain non-zero. We explicitly see this in part I of the
figure. On the other hand, when we consider time evolution from a
gapped phase towards a gapless phase, even though the initially
disconnected Fermi sea remains disconnected in phase space (again
working with the same evolution equation \eq{u-hermitian}), we find
that the projection on the $\t$-axis can become gapless, thus ensuring
the presence of a {\it gapped $\to$ gapless} transition.  

\paragraph{\label{para:GL}Relation to Gregory-Laflamme
transition:}
Note that the above argument is robust and ensures that the asymmetric
nature of dynamical transitions will not depend on the details of the
potential $V(U)$ in (\ref{unitary-action})(for example, it persists
for the double trace potential, as we will see in Section
\ref{sec:double-trace}). Now the choice of $V(U)$ dictates the rate at
which the phase space fluid is squeezed due to the dynamics of
individual particles; what happens when $V(U)$ is such that this
squeezing effect is very strong?  In this case, the neck will be
stretched thin and its thickness will shortly become $O(1/N)=
O(\hbar)$. At this stage, the semiclassical droplet description will
clearly be violated, and we will need to use the full quantum
mechanical description.  As discussed in Section \ref{sec:topology}, a
gap-opening transition in the double trace matrix model is related to
a Gregory-Laflamme transition in the dual string theory.  In a GL
transition, classical general relativity is not valid when the size of
the `neck' (of a black string or of a solitonic string) becomes Planck
size.  In case of the black string $\to$ black hole transition, the
breaking of the neck involves appearance of a naked singularity. In view
of the discussion above, this
singularity would be related to the size $O(1/N)$ neck region in the
matrix model.  Thus we expect that the understanding of matrix models
would be important to understand the issue of the Gregory-Laflamme
transition.  We will discuss this issue further in Section
\ref{sec:topology}.

\begin{figure}
\begin{center}
\includegraphics[scale=.5]{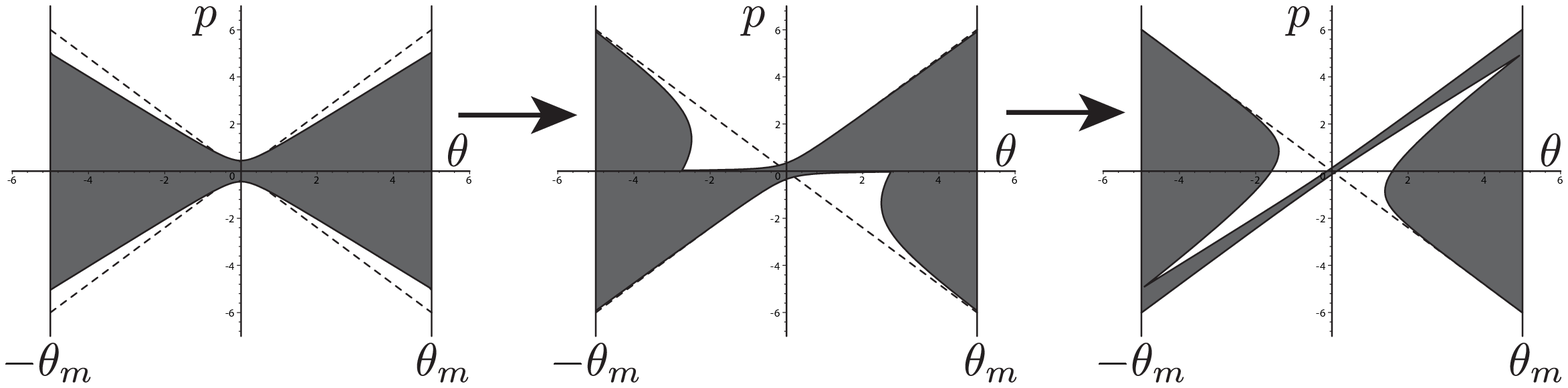}
\end{center}
\begin{center}
(I) $a_i < a_c $,  $a_f > a_c $.
\end{center}
\begin{center}
\includegraphics[scale=.5]{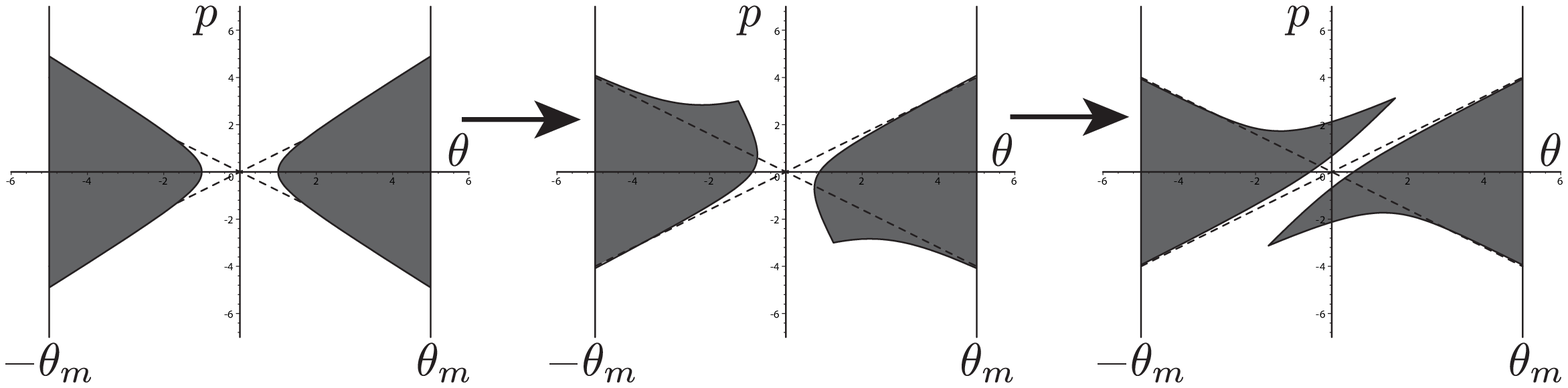}
\end{center}
\begin{center}
(II) $a_i > a_c $,  $a_f < a_c $.
\caption{\footnotesize Quantum quench dynamics of the filled Fermi sea  in the phase space   in the hermitian model \eq{hermitian}. 
  The dashed lines are $p=\pm \sqrt{a_f} \theta$, which are the separatrices for the single particle motion. 
   In (I), the left panel shows
  the shape of the initial filled Fermi sea, corresponding to the
  ground state for $a=a_i < a_c$. This is a gapless phase, hence the
  Fermi level is above the top and the droplet is a single connected
  one. The right panel shows an intermediate time-evolved configuration. After the quench, the modified hamiltonian stretches
  the droplet, but it cannot split it, hence the eigenvalue density,
  which is a certain projection on the horizontal axis (see
  \eq{collective-single}), also cannot split.  In (II), left panel,
  the initial $a=a_i> a_c$ is in the gapped phase. The Fermi level is
  below the top, hence the left and right `wells' are both filled up
  to the Fermi level, while remaining disconnected. Hence the filled
  Fermi sea has two disconnected segments (therefore, the eigenvalue
  distribution is gapped). The right panel shows the time-evolved
  configuration (according to $a=a_f < a_c$), which is such that
  although the droplet still consists of two disconnected parts in
  phase space, the projection on the horizontal axis is gapless, hence
  the eigenvalue density  is gapless.}
\label{fig:dyn-phase-tr}
\end{center}
\end{figure} 

\subsubsection{More on departure from adiabaticity}
\label{sec:departure}

In this subsection, we come back to the issue of the role
played by the quantum critical point in quantum quench dynamics (QQD). 
This has been 
discussed in several contexts: e.g. (a) how the
asymptotic density of excitations scales as a 
function of the parametric distance from criticality when the initial
hamiltonian is at criticality
(see \cite{Barankov:2009}), or (b)
how the asymptotic  QQD value of some observable or order parameter 
differs from the adiabatic value, as a function of (in our notation)
 $a_f- a_c$ 
(for given fixed $a_i$).  
In the above subsections, we found many examples of differences between
QQD and adiabatic evolution.  We explore
this aspect more in this subsection by studying the behaviour of
$\rho_1$.  We find that as $a_f$ is taken further away from $a_i$
towards the other side of the quantum critical point, the departure
from adiabaticity becomes appreciable from near $a_f \sim a_c$.  We
have found similar behaviour for the higher moments $\rho_2, \rho_3$
etc. as well.  This phenomenon of departure from adiabaticity around
the critical point is similar to that reported in \cite{Sengupta:2004}
in the context of QQD in a Hubbard model. Indeed our
Fig.~\ref{fig:departure-adiab} is quite similar to Fig.~3 of
\cite{Sengupta:2004}.\footnote{The departure from the adiabatic curve
  does not exactly coincide with criticality in \cite{Sengupta:2004}
  or in our work here, an effect which is more pronounced for higher
  moments.}  This points to some possible universal role played by the
quantum critical point even in dynamical phenomena.

\begin{figure}
\begin{center}
\begin{tabular}{cc}
\begin{minipage}{0.5\hsize}
\begin{center}
\includegraphics[scale=.4]{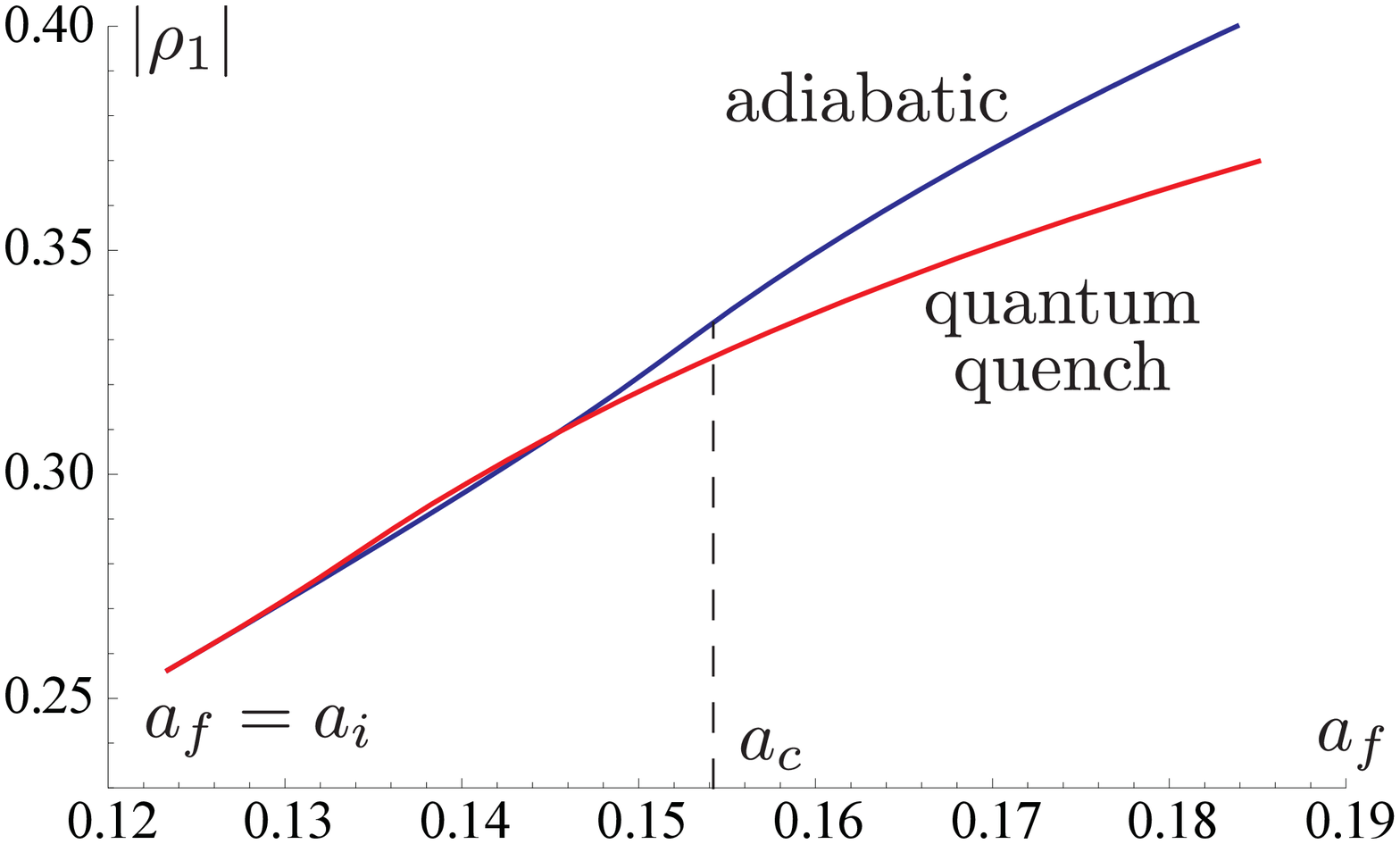}
\begin{center}
{\bf (a)} $a_i<a_c $
\end{center}
\end{center}
\end{minipage} 
\begin{minipage}{0.5\hsize}
\begin{center}
\includegraphics[scale=.4]{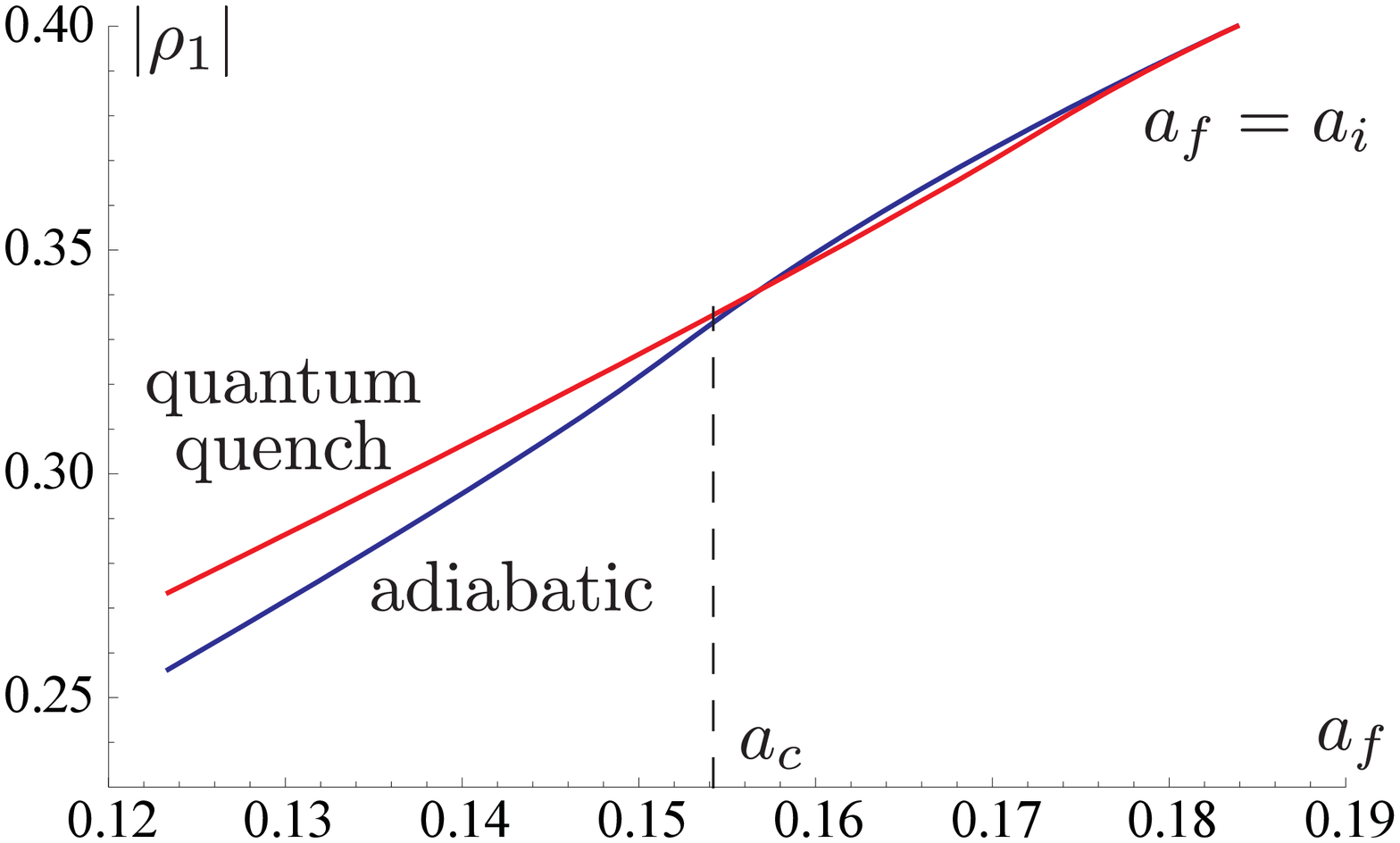}
\begin{center}
{\bf (b)} $a_i>a_c$
\end{center}
\end{center}
\end{minipage} 
\end{tabular} 
\caption{\footnotesize A plot of (the asymptotic value of) $\rho_1$ as
  a function of $a_f$. The setup is the same as that of Fig.~\ref{fig:no-gap}. In the left panel, we start with the gapless phase
  $a_f=a_i = 0.8 a_c$, and consider QQD experiments for progressively
  higher values of $a_f$ till we go across $a_f=a_c$ towards $a_f =
  1.2 a_c$. The blue curve shows the adiabatic value of
  $\rho_1$ (which corresponds to the ground state of the
  $a_f$-hamiltonian), while the red curve shows the large
  $t$ asymptotic value of $\rho_1(t)$, which agrees with the GGE result. The two curves start showing
  significant departure from around $a_f = a_c$. In the right panel,
  we start with the gapped phase with $a_f= a_i=1.2 a_c$ and end up
  with $a_f = 0.8 a_c$; the colour coding is similar. Once again we
  find that departure from adiabaticity starts from around
  $a_f=a_c$. This figure is similar to, e.g., Fig.~3 of
  \cite{Sengupta:2004}, where a similar pattern of departure from adiabaticity is
  reported for quantum quench in a Hubbard model.
   }
\label{fig:departure-adiab}
\end{center}
\end{figure}


\section{\label{sec:double-trace} Double trace model}

Like the single trace model, the double trace model \eq{double-trace},
can also be described (see \eq{fermion-double}) in terms of $N$
fermions moving in one spatial dimension. The hamiltonian is given by
\begin{align}
H &= \int d\theta~  \frac1{N^2} 
\del_\theta\psi^\dagger(\theta)
\del_\t\psi(\theta,t)+ H_{int},
\nn
H_{int} & = \frac\xi{N} \int dt~d\t~d\t'[\psi^\dagger(\t,t)\psi(\t,t)
\cos(\t- \t')\psi^\dagger(\t',t)\psi(\t',t)].
\label{hamiltonian-double}
\end{align}
The difference from the single trace case is that the fermions are now
interacting with a mutual $\xi\cos(\t-\t')$ potential.  Since this is
an interacting system, it is not obviously integrable; we will not
explore the question of integrability in this work.

Before going to explore dynamics in this system, we will briefly
review its connection to 2D adjoint scalar QCD and string theory.

\paragraph{\label{para:2d-qcd}The 2D adjoint scalar QCD:}

Consider the following two dimensional gauge theory on a $S^1$
\cite{Mandal:2011hb},
\begin{align} 
\kern-5pt S = \int \kern-5pt dt \int_0^{L}\kern-5pt dx \,
\Tr \Biggl( & \frac{1}{2{g}^2} F_{tx}^2 + \sum_{I=1}^D \frac12
\left(D_\mu Y^{I}\right)^2
+ \sum_{I,J}
\frac{{g'}^2}{4} [Y^I,Y^J][Y^I, Y^J] \Biggr).
\label{d=2}
\end{align}
Here $L$ is the period of the $S^1$.  This theory is related to a
dimensional reduction of $D+2$ dimensional pure Yang-Mills theory 
(which, for $D=8$, can be related to $N$ D2 branes compactified
on a Scherk-Schwarz circle)\footnote{If we derive
  the model (\ref{d=2}) from the higher dimensional gauge theory or D
  branes, additional terms are induced through quantum loop effects.
  For this reason, the coupling at the two dimensional gauge field $g$
  and the commutator interaction $g'$ have been distinguished in the
  model (\ref{d=2}).  The most relevant term in these loop corrections
  in weak coupling is the mass term for the adjoint scalar.  We omit
  this mass in this paper as argued in \cite{Mandal:2011hb}.}. 
  The  thermodynamical  phase structure of this model is as shown in Fig.~\ref{fig-phases-double}.
At sufficiently large $\beta$, (at least above the critical line $CBDO$ so that the system is  confined, $\Tr V=0$),
 the adjoint scalars would have a mass gap $\Delta$ and are integrated out through the $1/D$ expansion \cite{Mandal:2011hb} (see also
\cite{Aharony:2005ew, Semenoff:1996xg}). The theory is then described by the
following 
 effective action 
\begin{align}
S/DN^2= \int \kern-5pt dt \left[\frac{1}{2N} \Tr \left(| \del_tU|^2 \right)  - \frac{ \Delta}{ \pi \tilde{\lambda} L} \sum_{n=1}^{\infty}
\frac{
K_{1}(n \Delta L)}{n}
\left| \frac{1}{N}\Tr U^n \right|^2 
  \right],
  \label{effective-2d}
\end{align}
where $U(t) = P\exp[i\oint A_1(t) dx]$, and
$K_1$ is the modified Bessel function of the second kind.
In the following subsections, we would mainly consider dynamical phase transitions near the critical line  $AB$ in Fig.~\ref{fig-phases-double}.
In this case, we can ignore the higher modes ($n>1$) in the infinite sum in the potential term, and obtain the double-trace model \eq{double-trace} or \eq{hamiltonian-double}.
Thus the model \eq{hamiltonian-double} describes the dynamics of the spatial Polyakov loop $U(t)$ of the QCD (\ref{d=2}) in the confined phase.
Here $\xi$ in \eq{hamiltonian-double} is a function of the parameters in the QCD, and importantly it monotonically decreases as $L$ increases \cite{Mandal:2011hb}.

\paragraph{Phases:} 

The parameter $\xi$
determines the phase structure of the theory \eq{hamiltonian-double}. For high enough $\xi$
($\xi> \xi_2$), the attractive potential $S_{int}$ dominates and the
system is gapped. For low enough $\xi$ ($\xi< \xi_2)$, the kinetic
term dominates and the system is gapless. See Fig.~\ref{fig-phases-double} {\bf (b)} for more details.

\begin{figure}
\begin{center}
\begin{tabular}{cc}
\begin{minipage}{0.5\hsize}
\begin{center}
\includegraphics[scale=.7]{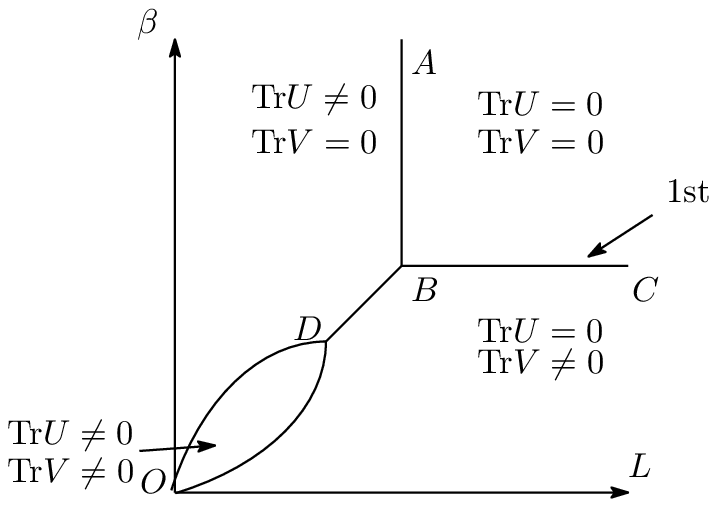}
\begin{center}
{\bf (a)} Phase structure of 2D adjoint scalar QCD (\ref{d=2}).
\end{center}
\end{center}
\end{minipage} 
\begin{minipage}{0.5\hsize}
\begin{center}
\includegraphics[scale=.25]{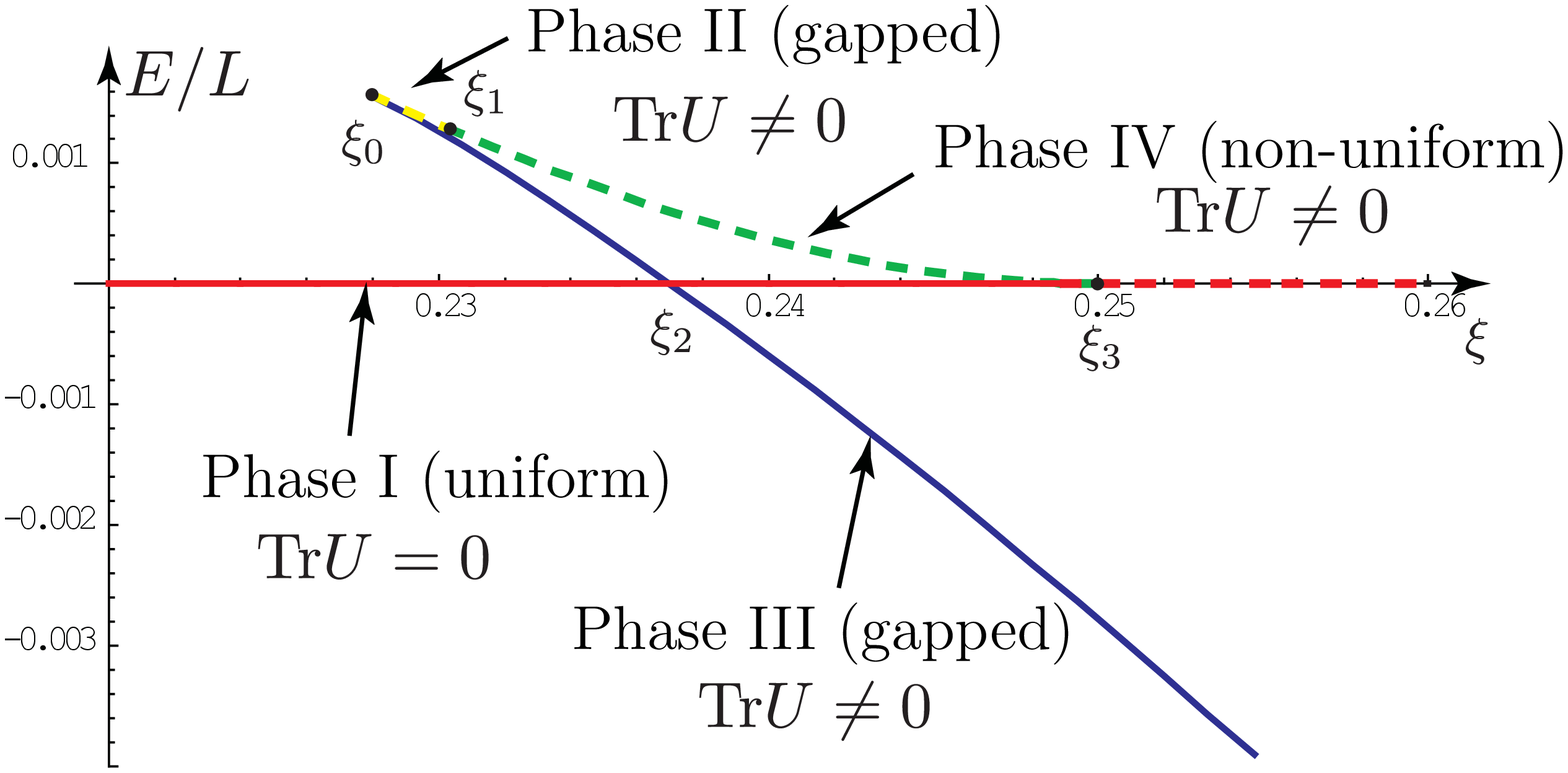}
\begin{center}
{\bf (b)} A plot of free energy around line $AB$ of diagram {\bf (a)}.
\end{center}
\end{center}
\end{minipage} 
\end{tabular} 
\caption{\footnotesize {\bf (a)}
Phases of the 2D QCD model \eq{d=2}
  are shown for various values of the inverse temperature $\b$ and
  spatial size $L$.
  The phases are characterised by the expectation values of the temporal Polyakov loop $V$ and the spatial Wilson loop $U$.
  We are interested in the limit of  zero temperature (non-compact time direction), in which the Polyakov
  loop $V=0$; the only phase boundary of interest here is the line
  $AB$, across which the spatial Wilson line $U$ changes.
  The theory in this region is described by
\eq{double-trace} or \eq{hamiltonian-double}. {\bf (b)} The energy $H=
E$ of \eq{hamiltonian-double}, computed using large $N$ Fermi liquid
picture, is plotted for various values of $\xi$ (see
\cite{Mandal:2011hb, Basu:2006mq} for details). To connect with diagram {\bf (a)}, note
that $\xi$ is a monotonically decreasing function of $L$.  Solid lines
represent the stable phase (the solution with the lowest value of $E$
for a given $\xi$) as well as metastable phases, while the dotted lines
represent unstable phases.  At $\xi_2= 0.237$ there is a 1st-order
transition from a stable uniform distribution to a stable gapped
one. Below $\xi_0=0.227$ the metastable gapped phase does not exist,
similarly above $\xi_3=1/4$ the metastable uniform phase does not
exist. At $\xi_1=0.231$ there is a Gross-Witten-Wadia (GWW) transition among two unstable phases.}
\label{fig-phases-double}
\end{center}
\end{figure}

\subsection{Dynamical evolution}
\label{sec-decay}

In case of the double trace model, several solutions have been derived for a given $\xi$ ($>\xi_0$) as shown in Fig.~\ref{fig-phases-double} {\bf (b)} but the ground state is always unique.
Hence we naively expect that the unstable or meta-stable solutions may evolve toward the ground state if we add sufficient perturbations to these solutions.
We will explore the question of possible dynamical phase transition as
well as the question of relaxation in this process.
In this section, we use the semiclassical approximation and consider only the large-$N$ case. (See the details in Appendix \ref{app:details}.)

\subsubsection{Dynamical evolution from the uniform gapless phase}
\label{sec-decay-gapless}

We will consider the time evolution of the uniform phase.
In terms of the droplet picture (Section \ref{app:droplet}), the
uniform phase is given by 
\begin{align}
\rho(\t)= 1/(2\pi), \quad {\cal P}=0.
\label{uniform-rho}
\end{align}
This configuration is always a solution of the model (\ref{hamiltonian-double}), since the force term vanishes for particles located
uniformly on the circle. 
In order to have non-trivial dynamics, we add a small perturbation to this configuration.  

Before we discuss the actual results of the evolution below, let us
pause and ask what to expect.
If $\xi$ in (\ref{hamiltonian-double}) was small enough, say well below $\xi_2= 0.237$, for such values of
$\xi$ the uniform phase is the ground state. 
Thus, we expect that for small enough amplitude of the
perturbation, the system will revert back to the
uniform phase. Indeed for $\xi= 0$, discussed in Section \ref{sec:cascade}, we find this to be explicitly true.

On the other hand, we are interested here in values of $\xi$
above $\xi_2$ where the ground state is the gapped one. 
Thus, we will expect the
perturbation  to be unstable towards formation of a
gapped phase.\\

\ni\underbar{Order parameter}: Since all moments \eq{moments} of the
density vanish in the uniform phase, any of these, say $\rho_1$, can be
considered as an order parameter, which would change from $\rho_1=0
\to \rho_1 \ne 0$ if the distribution changes from uniform to a
non-uniform solution.
  Note that we consider $Z_2$ symmetric ($\theta \leftrightarrow -\theta$) perturbations only in this article.
  Then we can take $\rho_1 = \Tr U /N$ by a gauge choice.
\\

We will now discuss the actual result of the time-evolution of
a perturbation according to \eq{particle-double-EOM}.  
In Fig.~\ref{fig-gapless-gapped-double}, the result is presented for
$\xi =0.260$, at which the uniform solution (\ref{uniform-rho}) is unstable.
Then $|\rho_1|$ develops to a non-zero value (see the right panel
of the figure), and we obtain a non-uniform distribution. 
However the distribution does not develop a gap, unlike the naive expectation above. 
Indeed, just as in Fig.~\ref{fig:dyn-phase-tr}, part (I), {\it the gapless $\to$ gapped transition is once again not possible since the droplet in phase space does not
  split}. 
  In addition to all this, we also find evidence for equilibration in $|\rho_1|$(see the right panel)
which indicates relaxation of the phase space configuration to one
with a thin stable neck.\footnote{The `neck'  avoids a gap-opening
transition, and consequently avoids a dynamical singularity. It is
important to explore this further, particularly in the context
of the double scaled matrix model. For similar issues related to
smoothening of singularities  through $1/N$ effects, please see
\cite{AlvarezGaume:2006jg,AlvarezGaume:2005fv}.}  

\begin{figure}[H]
\begin{center}

\begin{tabular}{ccc}
\begin{minipage}{0.35\hsize}
\begin{center}
\includegraphics[scale=.5]{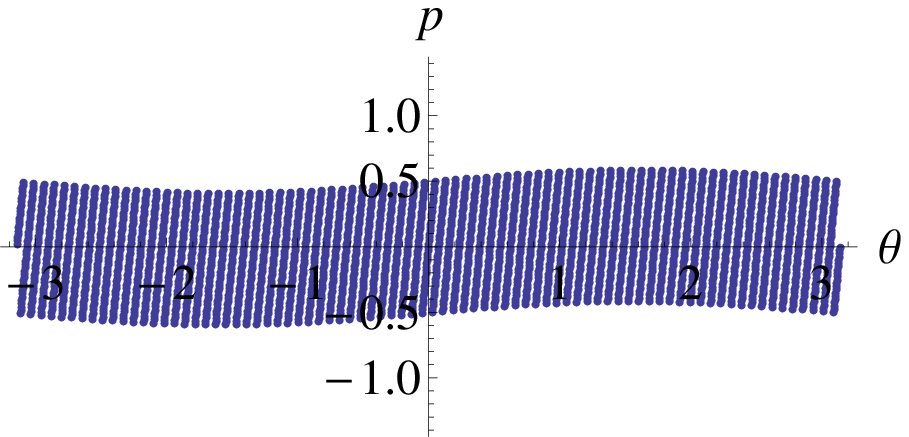}\\
$t=0$
\end{center}
\end{minipage} 
\begin{minipage}{0.35\hsize}
\begin{center}
\includegraphics[scale=.5]{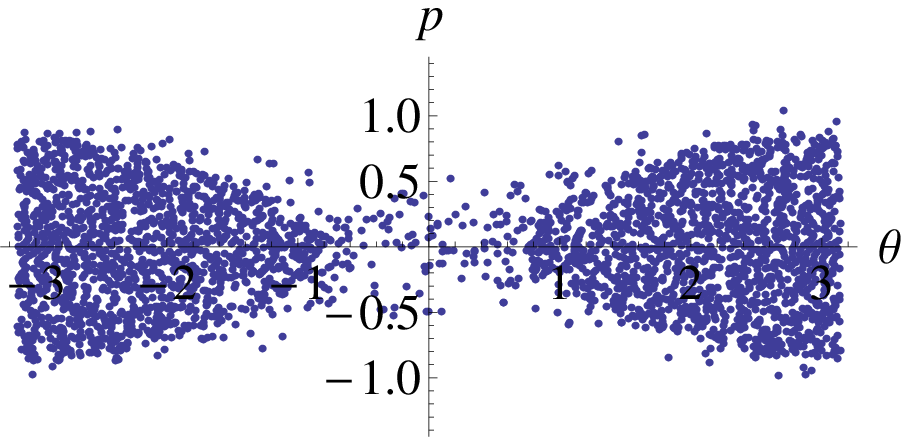}\\
$t=200$
\end{center}
\end{minipage} 
\begin{minipage}{0.3\hsize}
\begin{center}
\includegraphics[scale=.4]{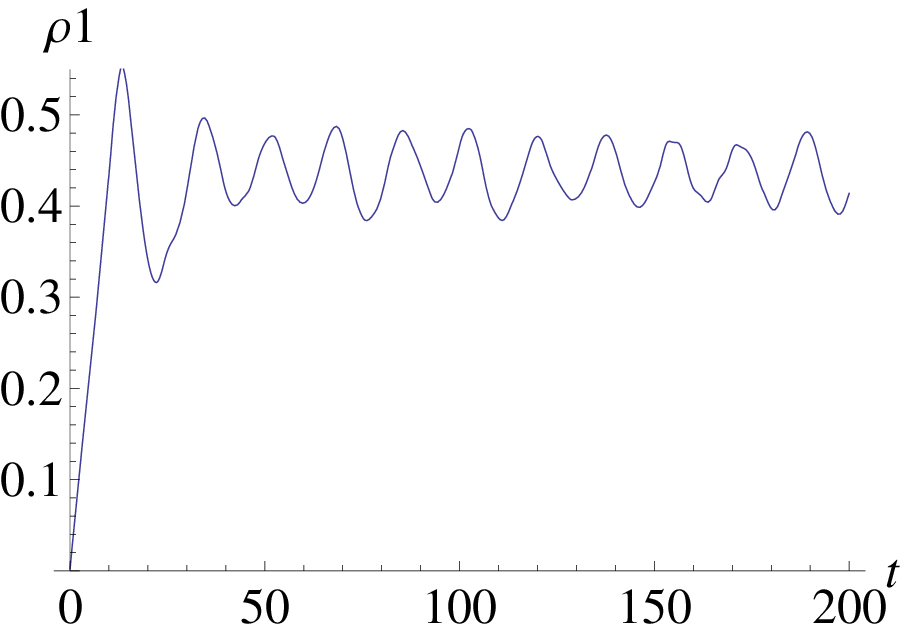}\\
$\rho_1(t)$
\end{center}
\end{minipage} 
\end{tabular} 
\caption{\footnotesize Time evolution of the uniform solution (\ref{uniform-rho}) with a small perturbation at $\xi=0.260$.
The left panel shows the initial configuration. 
The configuration approaches a gapped phase, but develops a thin neck,
  like in Part (I) of Fig.~\ref{fig:dyn-phase-tr} and the gap never
  develops.
  The right-most panels show the behaviour of the order parameter $\rho_1$ with time. It shows evidence of an equilibration.}
\label{fig-gapless-gapped-double}
\end{center}
\end{figure}

\subsubsection{Time evolution from the gapped phase}

The discussion is analogous to the previous case. 
We consider a meta-stable gapped state at $\xi < \xi_2$ (Phase III in Fig.~\ref{fig-phases-double} {\bf (b)}) 
and perturb the system.

The result at $\xi=0.230$ is shown in Fig.~\ref{fig-gapped-other-double}.
A dynamical {\it gapped $\to$ gapless transition takes place}.
Thus, like in the single trace case, we once again see a one-way
dynamical phase transition, {\it gapped to gapless, but not gapless to
  gapped}. This once again, has a clear interpretation in terms of
phase space.

Note that if the perturbation is not sufficient, the state reverts back to the original gapped configuration, since the configuration is meta-stable.

\begin{figure}[H]
\begin{center}
\begin{tabular}{ccc}
\begin{minipage}{0.35\hsize}
\begin{center}
\includegraphics[scale=.5]{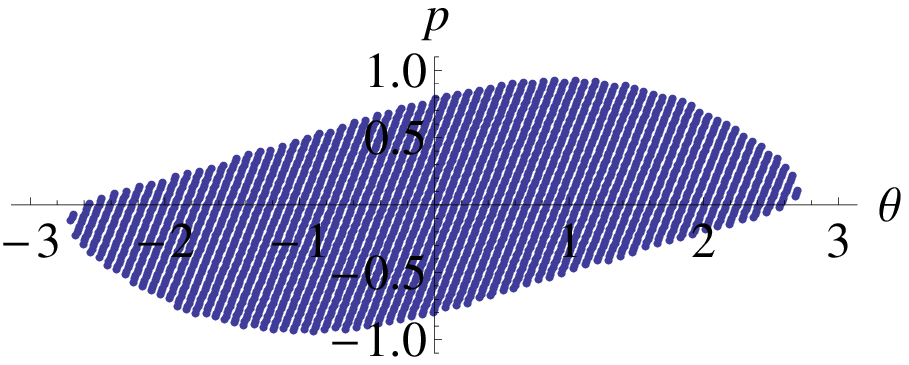}\\
$t=0$
\end{center}
\end{minipage} 
\begin{minipage}{0.35\hsize}
\begin{center}
\includegraphics[scale=.5]{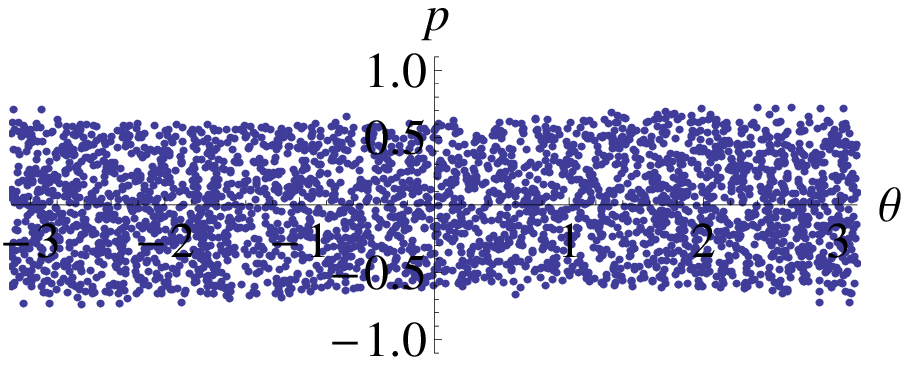}\\
$t=400$
\end{center}
\end{minipage} 
\begin{minipage}{0.3\hsize}
\begin{center}
\includegraphics[scale=.4]{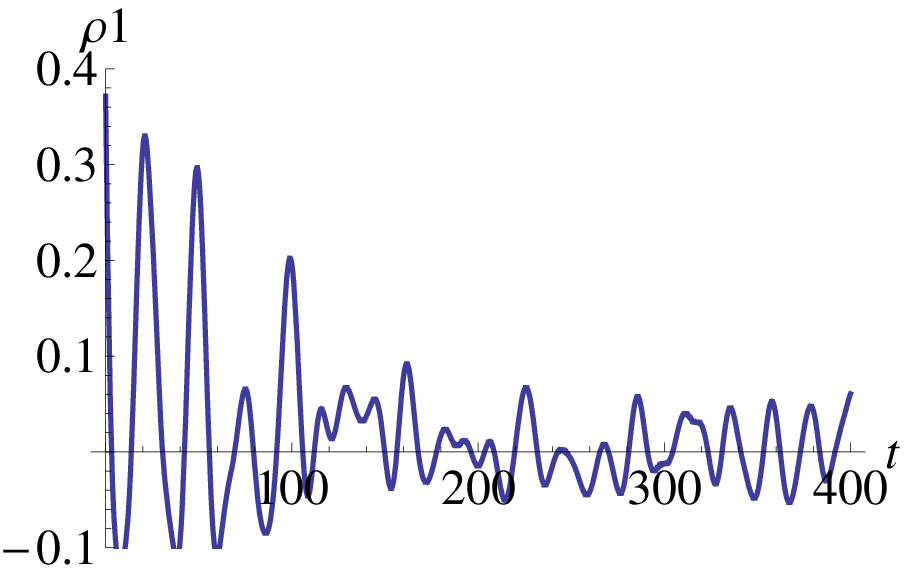}\\
$\rho_1(t)$
\end{center}
\end{minipage} 
\end{tabular} 
\caption{\footnotesize 
 The figure on the left shows a slightly perturbed meta-stable gapped distribution at $t=0$. 
 The value of $\xi$ is 0.23. 
 The figure in the middle shows a gapless distribution at $t=400$.  
 The figure on the right depicts $\rho_1(t)$ as
  it changes from 0.4 at $t=0$ to 0.}
\label{fig-gapped-other-double}
\end{center}
\end{figure}

\subsubsection{Appearance of exotic states}

We have considered the time evolutions of unstable/meta-stable solutions by adding perturbations.
However if we add sufficiently large perturbations, we observe appearance of exotic asymptotic states.
See Fig.~\ref{fig-two-solitons}, e.g., where in the middle panel
two blobs of fermions have appeared, which are moving towards the left and the
right respectively.  As time
progresses, the two blobs keep executing a period motion and
in coordinate space, this means that two density peaks approach,
merge and separate out, and repeatedly pass through each other 
like colliding solitons (remember
that in our model the fermions are moving in a circle).

\begin{figure}[H]
\begin{center}
\begin{tabular}{ccc}
\begin{minipage}{0.35\hsize}
\begin{center}
\includegraphics[scale=.5]{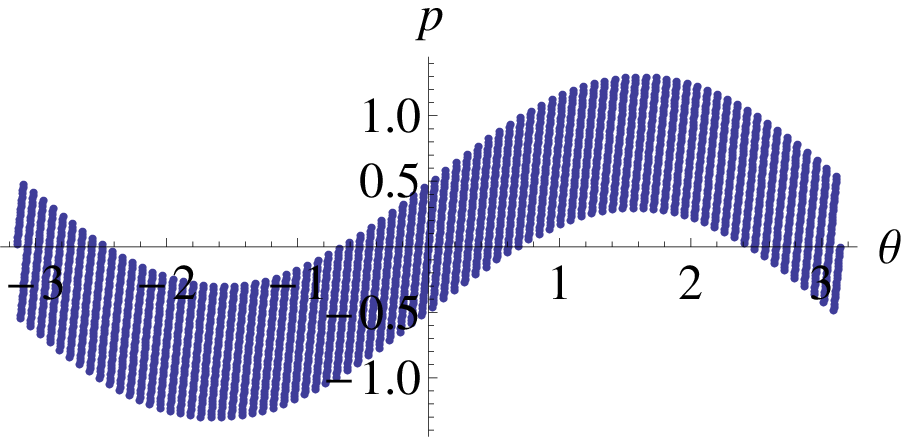}\\
$t=0$
\end{center}
\end{minipage} 
\begin{minipage}{0.35\hsize}
\begin{center}
\includegraphics[scale=.5]{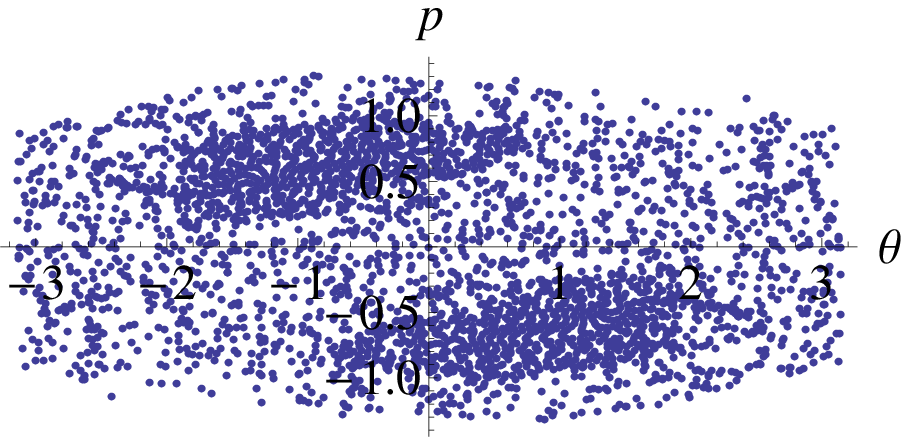}\\
$t=200$
\end{center}
\end{minipage} 
\begin{minipage}{0.3\hsize}
\begin{center}
\includegraphics[scale=.4]{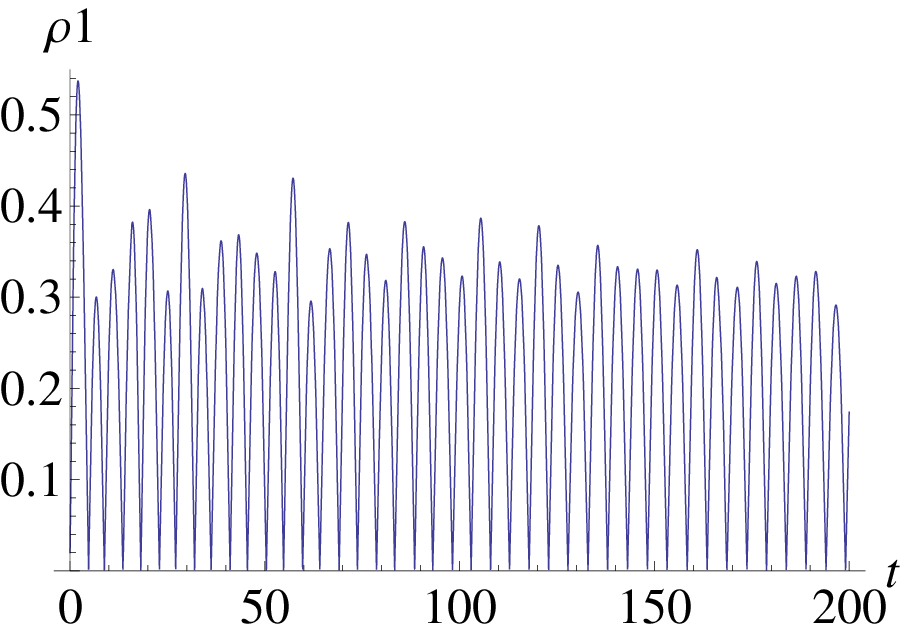}\\
$\rho_1(t)$
\end{center}
\end{minipage} 
\end{tabular} 
\caption{\footnotesize Time evolution of the unstable uniform solution
  (\ref{uniform-rho}) with a large perturbation at $\xi=0.260$.  The
  particles split into two well-defined blobs in phase space. The
  upper blob moves to the right while the lower blob moves to the
  left, and they keep moving past each other periodically. In
  coordinate space, the blobs overlap; this implies that two density peaks
  periodically pass through each other like colliding solitons.}
\label{fig-two-solitons}
\end{center}
\end{figure}

Another exotic time evolution can be seen if we consider the higher modes of the potential (\ref{effective-2d}).
We assume appropriate values of $\Delta$ and  small $L$ such that $\xi$ is large, consequently the gapped solution is stable and the uniform one (\ref{uniform-rho}) is unstable.
In this setup, the unstable uniform solution (\ref{uniform-rho}) evolves as shown in Fig.~\ref{fig-2peak}.
In this case, the dynamical phase transition consists of two steps.
First the unstable uniform state decays to 
a 2-peak state (where the `peaks' refer to those in the coordinate space density $\rho(\t))$.
This  2-peak state remains for some time, and finally decays to the one-peak state.
We have also observed appearance of various multi-peak intermediate states if we change the initial perturbations.

Indeed the existence of unstable multi-peak solutions for the small $L$ region in Fig.~\ref{fig-phases-double} {\bf (a)} and the possibility of their appearance during the decays of the unstable uniform solution have been predicted in \cite{Azuma:2012uc}.
Our result confirms this conjecture.
Interestingly similar time evolutions have been observed in the decay of black string in the Gregory-Laflamme transition also \cite{Choptuik:2003qd, Lehner:2010pn}.
An unstable black string does not always decay directly to a single-black hole, which is thermodynamically most stable, but to multiple black holes in some cases.

\begin{figure}[H]
\begin{center}

\begin{tabular}{ccc}
\begin{minipage}{0.35\hsize}
\begin{center}
\includegraphics[scale=.5]{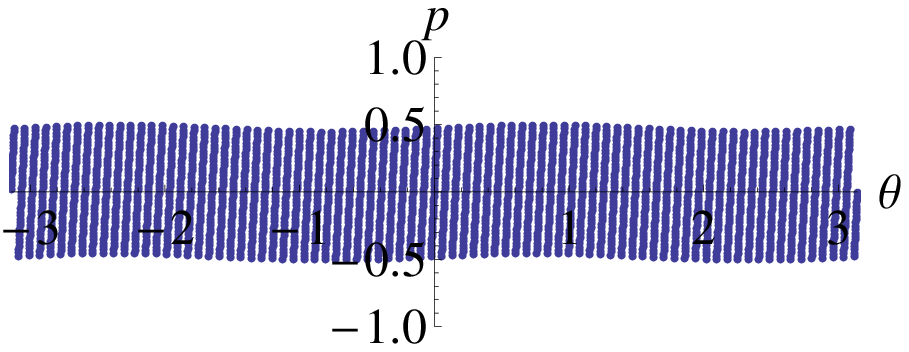}\\
$t=0$
\end{center}
\end{minipage} 
\begin{minipage}{0.35\hsize}
\begin{center}
\includegraphics[scale=.5]{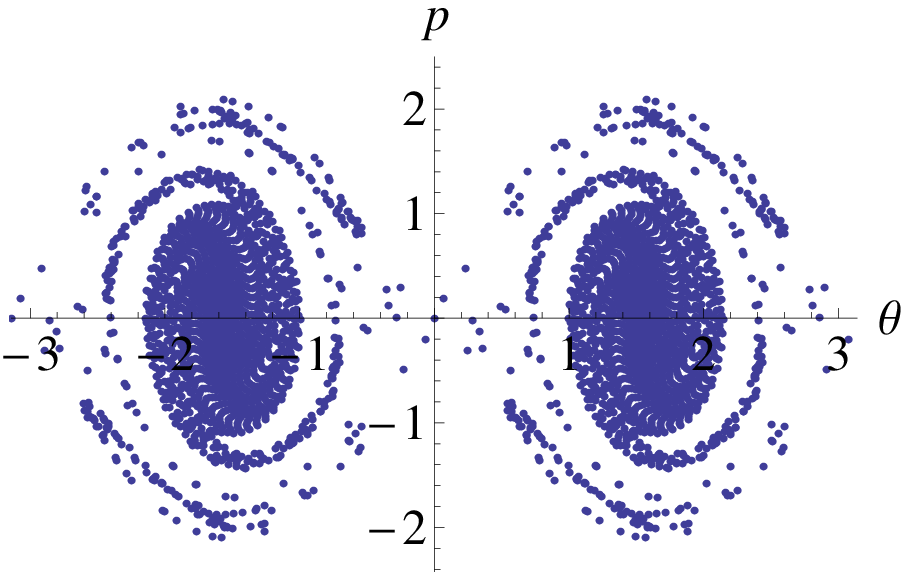}\\
$t=10.0$
\end{center}
\end{minipage} 
\begin{minipage}{0.3\hsize}
\begin{center}
\includegraphics[scale=.35]{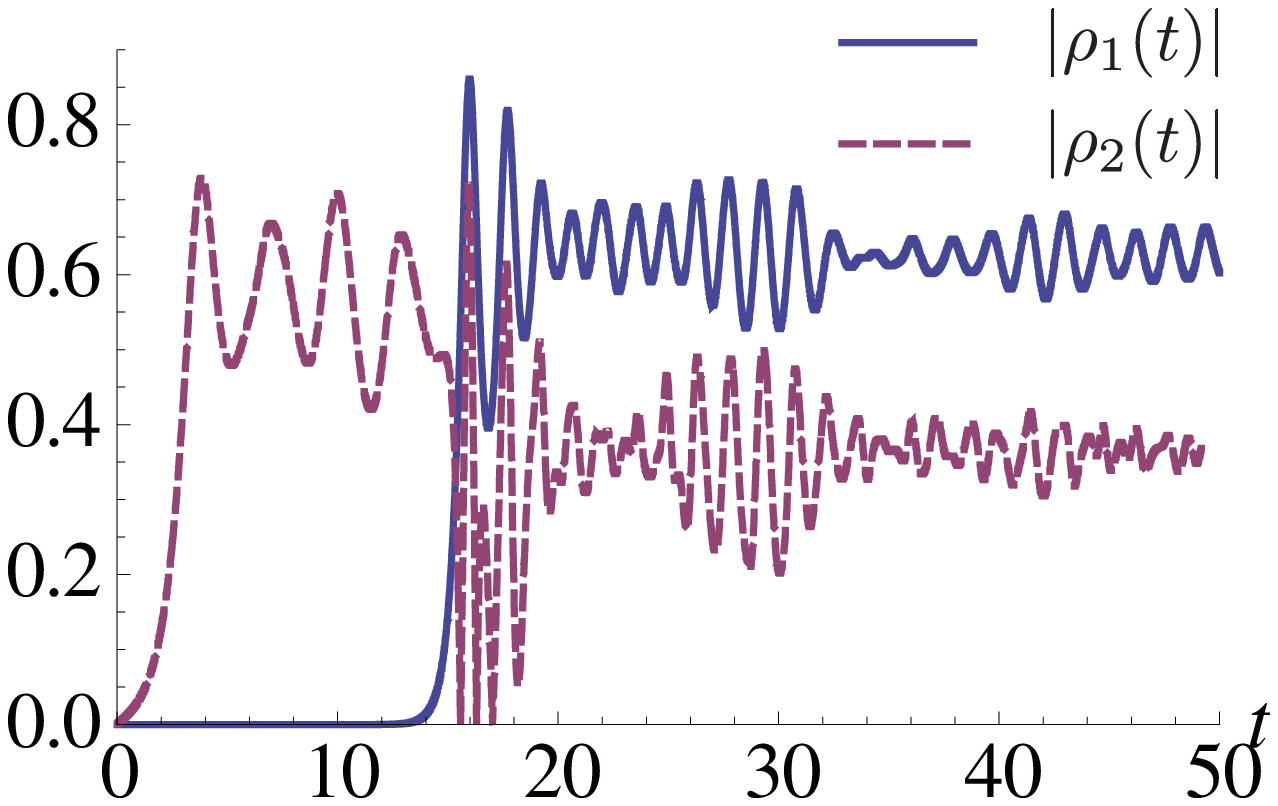}\\
$\rho_1(t),~ \rho_2(t)$
\end{center}
\end{minipage} 
\end{tabular} 
\caption{\footnotesize 
The appearance of the 2 peak state during the time evolution of the uniform configuration.
We take $\xi=5.0$ and $\Delta L = 1.0$.
The strong signal of $\rho_2$ indicates the two peaks in the eigenvalue.
First the uniform state decays to the 2-peak state around $t=1$, and then it decays to one peak state around $t=15$.
(We do not plot the one peak solution here but it is similar to Fig.~\ref{fig-gapless-gapped-double}).
}
\label{fig-2peak}
\end{center}
\end{figure}

Note that the two-peak state that appears in Fig.~\ref{fig-2peak} is qualitatively different from the breather-like asymptotic state in Fig.~\ref{fig-two-solitons}.
The former state is approximately static (with two
distinct, long-lived density peaks around $\t=\pm\pi/2$) until it decays
to a single-peak state, whereas
the latter state is approximately stationary and the two peaks merge and
separate out repeatedly in an oscillatory fashion.
(See the movies for these evolutions from the link in footnote \ref{ftnt-movie}.)

As we have seen in section \ref{sec:single-trace}, the system is integrable if the double trace interactions are turned off.
One open question is whether this interaction breaks the integrability.
The appearances of these exotic asymptotic/intermediate states may shed light on this issue.
Especially the exotic asymptotic states in Fig.~\ref{fig-two-solitons} may imply an attractor structure in this system.
We will come back to this issue in a future work.

\section{\label{sec:discussions}Discussions}

The results of our paper have already been summarised in Section
\ref{sec:intro} and described in Sections \ref{sec:single-trace} and
\ref{sec:double-trace}. In this section, we will describe some related
issues, speculations and outstanding questions.

\subsection{Duality to 2D string theory and $W_\infty$ algebra}
\label{sec:c=1}

The model defined by Eqs.~\eq{hermitian}, for fixed $a$, is
essentially the $c=1$ matrix model. This model (in its double-scaled
form: $(a- a_c)N = \mu$), has a string theory `dual', namely the
two-dimensional bosonic string theory, which is characterized by a
massless scalar field (called `tachyon') on a flat spacetime and
linear dilaton background (see, e.g. the review of
\cite{Ginsparg:1993is}). The tachyon field fluctuation is given by a
non-linear and non-local transformation of the matrix model scalar
field fluctuation (see, e.g., \cite{Natsuume:1994sp, Dhar:1995nq,
  Dhar:1995np} for more details). The filled Fermi sea, $\rho(\t)=
\frac1\pi \sqrt{2(\mu-V(\t))}$, ${\cal P}=0$, is a solution of
\eq{rho-EOM}, and corresponds to the tachyon vacuum in string theory.
In quantum quench dynamics, in the matrix model description, we start
with an unstable density perturbation, which evolves in time under the
restoring force towards the above-mentioned stable configuration.
Since the motion is conservative (in fact, it has an infinite number
of conserved quantities), the configuration cannot reach the stable
configuration; in stead there is an oscillatory motion, in which lower
frequency modes transfer energy to the higher frequency ones, leading
to an energy cascade as discussed in Section
\ref{sec:cascade}.\footnote{We show this in the unitary matrix model,
  but we find similar behaviour also in the Hermitian matrix model
  which is directly identified as the $c=1$ matrix model.} 

It would be interesting to map this oscillation to string theory
tachyon modes by applying the transformation alluded to above. We
expect that qualitatively it would be a similar phenomenon to the one
found in the matrix model, with a finite energy extended configuration
giving rise to proliferation of high frequency modes, asymptotically
approximated by a general Gibbs ensemble (GGE) in spacetime. Note that
in the 2D string theory too, there are an infinite number of conserved
charges, which are the diagonal elements of an infinite dimensional
algebra, {\it viz.} the $W_\infty$ algebra (see \cite{Das:1991qb}),
thus making the GGE a natural choice of ensemble. It is an interesting
question whether the GGE can be represented by a new field
configuration such as a non-trivial solitonic configuration, or a
black hole.\footnote{It has been argued that inflaton oscillations
  around its vacuum configuration, which is analogous to the tachyon
  oscillations described above, can produce primordial black holes
  (see \cite{Carr:2005zd}, for example).} We discuss the black hole
scenario in some detail in the subsection \ref{sec:bh} below.

In \cite{Douglas:2003up}, a different, type 0, 2D string theory is
proposed as dual to the (gauged) hermitian model \eq{hermitian}, where
the singlet condition is ensured by a gauge covariant kinetic term
$\Tr(\del_t + i [A_0, M])^2$. In the context of the type 0B theory,
there are two massless scalars, the tachyon and an axion. We expect
many of the comments in the previous paragraph to go through in this
case as well.

\subsection{\label{sec:others}Gravity duals of other integrable systems}

The phenomena of selective equilibration and emergence of GGE studied
in this paper would be of interest in many other examples of
integrable systems with gravity duals. We mention below a few of them.

\paragraph{D1-D5 system:}\label{para:d1-d5} The D1-D5 system (reviewed in, e.g. 
\cite{David:2002wn}) is described, at a certain point $P$ of its
moduli space, by a free 2D CFT (with target space an orbifold of the
type $M^N/S(N)$, where $M= T^4$ or $K3$, and $N=N_1 N_5$). The usual
D1-D5 supergravity solution is located at a different point $Q$ of the
moduli space, which can be reached by introducing a marginal
interaction $\lambda \int O$ to the free CFT at $P$. There are certain
nonrenormalization theorems which ensure the protection, under the
$\l$-deformation, of a number of quantities including black hole
entropy and absorption cross-section. In \cite{Das:2008ka} a certain
response function in the $\l=0$ integrable theory is exactly computed
and shown to agree with its gravity dual, namely the absorption
cross-section of the two-charge black hole (at the $\l=\l_0$
supergravity point). The agreement is found only in the large $N$
limit and using a coarse-grained probe. The field theory calculation
seems to be well-suited for a GGE description.

\paragraph{Giant gravitons:} The half-BPS configurations of
${\cal N}=4$ SYM on $S^3 \times R$ are described by a complex matrix
QM in a harmonic oscillator potential which can be described by free
fermions \cite{Mandal:2005wv, Takayama:2005yq} in a manner similar to
that described in this paper. The gravity duals are the well-known LLM
geometries \cite{Lin:2004nb}, which can be described in terms of giant
gravitons \cite{McGreevy:2000cw, Grisaru:2000zn, Hashimoto:2000zp} in
the probe approximation.

\paragraph{Higher spin duals of 2D CFT's:} Recently, \cite{Gaberdiel:2010pz}
has proposed a duality between the large $N$ limit of certain 2D CFT's
called the $W_N$ minimal models (represented by the coset WZW models
$SU(N)_k \times SU(N)_1/SU(N)_{k+1}$) and a one-parameter higher spin
(Vasiliev) theory parametrized by $\l= N/(k+N) {\rm fixed}, k,N \to
\infty$ (see, e.g. the review \cite{Gaberdiel:2012uj}). The $\l=0$ limit
corresponds to $N$ 2D relativistic free (complex) fermions and is very
similar to the system described in this paper.

\paragraph{\label{para:2d-gauge-theory}D2 branes:} The gauge theory system 
\eq{d=2} was considered in \cite{Mandal:2011hb}, where a gravity dual
is discussed (see Section 5 of that paper) in terms of the near
horizon geometry of D2 branes compactified on a `Scherk-Schwarz
circle' (i.e. a small spatial circle of size $L_2$ with antiperiodic
fermions). The other two directions are periodic, with sizes $\b, L$,
which characterize the phase diagram of the model is given in Fig.~\ref{fig-phases-double} {\bf (a)}. As discussed in \cite{Mandal:2011hb}, the
low temperature phase $\b > \b_{cr}$ of the model is described by a
soliton solution wrapping the temporal circle uniformly, while the
high temperature (`deconfinement') phase, $\b < \b_{cr}$, is given by a
localized soliton. We clearly find evidence of equilibration in the
low temperature phase of the field theory in
Fig.~\ref{fig-gapless-gapped-double} {\bf (a)}. It is also known that
the deconfinement phase shows dissipation. It would be interesting to
understand such equilibration in terms of the solitonic geometries
which obviously do not have a horizon.

\paragraph{\label{para:N=4-SYM}${\cal N}=4$ super Yang Mills theory:}
This is clearly one of the most important examples of integrable models
in AdS/CFT \cite{Beisert:2010jr}. The giant gravitons described above
describe a special sector of this model. More generally, operators of
${\cal N}=4$ super Yang Mills theory whose energies do not scale with
$N$, can be mapped to states in integrable spin chains which possess
an infinite number of conserved charges. For energies $E \sim O(N^2)$
the integrability does not persist; thus, a generic perturbation
equilibrates, and is described in the gravity dual by black hole
formation (see, e.g., \cite{Bhattacharyya:2009uu}). The present
discussion opens the possibility of equilibration of certain
perturbations in the integrable subsector and emergence of a GGE;
it would clearly be of interest to find a gravitational dual of 
such an ensemble, if any.

\subsection{\label{sec:bh}Black holes and $O(N)$ entropy}

It is tempting to speculate that the equilibration, discussed above
for the matrix QM model, and possibly generalizable to other
integrable field theories, is described in terms of formation of black
holes in a dual theory. In many known examples of AdS/CFT,
equilibration does have a natural interpretation in terms of formation
of black holes\footnote{or, in the context of a probe approximation,
  in terms of the damping of quasinormal fluctuations in a black hole
  background.}.

Such a scenario, however, must pass the following stringent
tests:\\ (i) Black holes are typically not associated with integrable
models; while the latter have an infinite number of conserved charges,
black holes typically have only a few ({\it cf.} uniqueness
theorems). Since a GGE \eq{GGE} is characterized by an infinite number
of chemical potentials, any black hole describing a GGE must also have
an infinite number of parameters. \\(ii) The entropy of the GGE must
match the entropy of the known black holes.\\ (iii) Besides the static
black hole solution, there must be a dynamical collapse solution which
would capture the equilibration process. The black hole formation time
should then match the time of relaxation to equilibrium described by
the GGE. \\ (iv) A non-extremal black hole will Hawking radiate; there
must be an interpretation of the two-stage process--- formation and
evaporation--- in the field theory description. \\ (v) In case we seek
a description of equilibration and dissipation in terms of absorption
of energy by a fixed black hole background in the dual geometry.

\paragraph{2D:} As mentioned above, the $c=1$ matrix model
\eq{hermitian}, is dual to the 2D bosonic string theory, in the linear
dilaton background. This theory, as described by its massless fields
(the metric, dilaton and the massless `tachyon'), does have the MSWW
black hole solution \cite{Mandal:1991tz, Witten:1991yr} (see
\cite{Gukov:2003yp,Berkovits:2001tg} for superstring
generalizations).  Let us discuss the above points to see whether this
black hole can be relevant for a description of the GGE found in the
present paper. \\ Point (i): the MSWW black hole solution, is indeed
the unique solution of the effective field theory of massless modes of
2D string theory, and is specified by only the mass
parameter \footnote{The charged cousins described in
  \cite{Gukov:2003yp, Nappi:1992as} require fluxes which can come from
  superstring or type 0 theories}. However, this solution can be
extended to an infinite-parameter generalization in 2D string field
theory \cite{Mukherji:1991kz}. Each such black hole possesses an
infinite number of specified $W_\infty$ charges (corresponding to
charges of higher mass non-propagating gauge fields), and could
potentially correspond to a given GGE with fixed values of the
chemical potentials. \\ Point (ii): the entropy of the MSWW black hole
was found to be (see \cite{Nappi:1992as}) of the order of $1/g_{str}^2
\sim N^2$, whereas the entropy of the GGE in Section \ref{sec:entropy}
turns out to be of order $N$.  \\ Point (iii):
\cite{Karczmarek:2004bw} has argued that if we wish to form the MSWW
black hole (or its type 0B counterpart) by collapsing a tachyon shell,
physics becomes strongly coupled even before a horizon is formed. In
the type 0B black hole, one can in stead consider collapse of an axion
shell; in this case, there is no strong coupling problem but the shell
is scattered back to infinity from near the incipient horizon and a
black hole is not formed.  \\ Point (iv):  Reference
\cite{Karczmarek:2004bw} performs a matrix model
computation to rule out any Hawking radiation from a possible gravitational  
collapse of axionic matter.  \\ Point (v): a fixed black hole
background in type 0A theory was argued in \cite{Gukov:2003yp} to be
dual to a deformed matrix model \cite{Jevicki:1993zg}. One could in
principle consider fermionizing such a model (as in \eq{fermion-single}), explore
dissipation in this, and attempt to explain this in terms of black
hole absorption. However, doubts have been raised in
\cite{Davis:2004xb} about the duality, as the entropy of the matrix
model is computed there to be $O(1)$ whereas the entropy of the black
hole is $O(N^2)$. The suggested dual matrix model, according to
\cite{Davis:2004xb}, which reproduces an $O(N^2)$ entropy, is the one
proposed in \cite{Kazakov:2000pm}; however, in this model, the
off-diagonal matrix elements play a crucial role, and the model is not
(at least in any obvious sense) integrable any more. One
could also consider the proposal in \cite{Dhar:1995nq} for construction of
a black hole background in terms of an asymmetrically filled Fermi sea.
We leave further studies of these issues for the future.

\paragraph{Higher spin black holes in AdS$_3$:} We have already
discussed above the duality between large $N$ 2D CFT's ($W_N$ minimal
models) and higher spin (Vasiliev) theories in AdS$_3$. In the limit
of $\l=0$, the field theory is that of $N$ number of complex free
fermions, and is likely to equilibrate as in this paper, with
emergence of GGE. As in case of the 2D gravity duals, the AdS$_3$ also
has black holes, the most well-known being the
BTZ black hole. Indeed, it has been found in \cite{Kraus:2011ds}
that the BTZ black hole has spin three and spin four generalizations. 
It would be very interesting to investigate if this result
generalizes to arbitrarily high spin, in which case they would
be natural candidates for a dual description of GGE.
These black holes have an entropy of $O(N)$, matching the
central charge $c\sim N$ of the field theory (see \cite{Ammon:2012wc},
e.g.), which agree with the $O(N)$ entropy of the GGE computed in this
paper. 

\paragraph{The superstar geometry:} In the context of the 
matrix model-LLM-giant graviton duality mentioned earlier, thinned out
phase space configurations such as \eq{u-val-coll} in which the
fermion occupation numbers are fractional have been studied. The
resulting LLM geometries are singular and are called superstar
geometries. These geometries have zero horizon area, but a Bekenstein
entropy can be assigned to a stretched horizon
\cite{Suryanarayana:2004ig}, which turns out to be $O(N)$ just like
the standard Boltzman formula for entropy in these systems, and
agrees with the GGE estimate of entropy. 

\paragraph{D1-D5:} The possible relevance of GGE to the two-charge 
D1-D5 black hole is already mentioned in the previous subsection. 
The entropy of the two-charge black hole $S \sim \sqrt{Q_1 Q_5}
\sim 1/g_{str}$ appears to agree with the $O(N)$ entropy of GGE.

\subsection{\label{sec:topology}Topology change and the
Gregory-Laflamme phase transition} 

The uniform and gapped phases of the double trace model
\eq{double-trace} correspond to the large $L$ and small $L$ sides of
the phase boundary AB in Fig.~\ref{fig-phases-double} {\bf (a)}. These
phases, by definition, refer to the uniform and clumped eigenvalue
distributions of the spatial Wilson line $U$, which arises from the
gauge of theory of D2 branes wrapped on a small Scherk-Schwarz
circle. (see page \pageref{para:2d-gauge-theory}). What do these
phases mean in gravity?  This question was investigated in detail in
\cite{Azuma:2012uc, Mandal:2009vz, Mandal:2011ws, Morita:2012hd} (see
\cite{Aharony:2005ew,Aharony:2004ig} for other related references).
The Euclidean near-horizon description of such a system at large $\b$
and $L$ (sizes of the temporal and spatial circles, respectively) is
that of a D2 soliton. If we T-dualize this system along the large
spatial circle, we get a D1-brane soliton which is uniformly smeared
along a small dual circle $L' \propto 1/L$. This is the description of
the phase to the right of the line AB in Fig.~\ref{fig-phases-double} {\bf (a)}; now as we decrease $L$, the dual circle
$L'$ starts to increase and the uniformly smeared D1 soliton starts
getting stretched and as $L'$ increases beyond a certain critical
size, it becomes unstable towards breaking and forming a localized
soliton.  In \cite{Gregory:1993vy}, Gregory and Laflamme predicted
such a transition as a thermodynamic phase transition\cite{Gregory:1994bj, Hovdebo:2006jy, Miyamoto:2007mh, Frolov:2009jr}. Thus, the
dynamical gapless $\to$ gapped transition in our model translates to a
dynamical GL transition from a uniform to a localized soliton.  In
particular, the observed non-existence of such a transition in the
matrix model, ultimately due to non-splitting of Fermi liquid
droplets, implies impossibility of a dynamical GL transition
from a 1-brane (uniform soliton) to a 0-brane (split solution). 

The original context of the GL transition was in the context
of a clumping transition of a black string into a black hole.
The question about whether this can happen dynamically is
related to the issue of cosmic censorship and the appearance of a
naked singularity, and has been discussed extensively in the
literature \cite{Choptuik:2003qd, Lehner:2010pn, Kol:2004ww, Horowitz:2001cz, Marolf:2005vn,
   Garfinkle:2004em}. In a supersymmetric
variant of \eq{d=2}, the phase transition discussed in the
previous paragraph would correspond to such a transition
from a black string to a black hole. 
Since the technique
developed in this paper appears to be fairly robust, dynamical
transition in this model could share similar features,
and could shed light on the issue of dynamical appearance
of a naked singularity. See also the related discussion in
page \pageref{para:GL}. 

 \subsection{\label{sec:confined-geom}
Non-ergodic time evolution in confinement geometries } 
Very recently, non-linear instabilities of global AdS geometries under perturbations are actively studied \cite{Bizon:2011gg, Jalmuzna:2011qw, Dias:2011ss, Buchel:2012uh}.
There, depending on initial perturbations, the
perturbed geometries evolve to small black holes, boson stars, 
or geons \cite{Dias:2012tq}.
 Since the final states depend on initial conditions,  the time evolutions are non-ergodic.
Since global AdS would correspond to a confinement phase in the dual gauge theory, the appearance of these non-ergodic evolutions sounds reasonable.
Especially, they would be related to the non-ergodic time evolutions in the D2 brane model (\ref{d=2}) in the confinement phase, as argued in section \ref{sec:double-trace}.
Since the gravity dual of the ground state
(which is in confinement phase), is a soliton geometry, as
shown in \cite{Mandal:2011hb} based on the Witten's holographic QCD,
we presume that the non-ergodic time evolutions would be observed in this soliton geometry too.
Then we can directly compare the time evolutions in the confinement phase in the gauge theory and the time evolutions in the soliton geometry.
This would be an interesting new direction of the gauge/gravity correspondence, since we can calculate both explicitly.

\subsection*{Acknowledgement}
We would like to thank Sumit Das, Kedar Damle, Avinash Dhar, Rajesh
Gopakumar, Sourendu Gupta, Hideo Kodama, Hong Liu, Subhabrata Majumdar, 
Shiraz Minwalla, Krishnendu Sengupta, Spenta Wadia and Xi Yin for 
discussions. 
We also thank Pallab Basu, Manavendra Mahato and Spenta Wadia for
collaboration on some earlier calculations which have overlaps with 
Section \ref{sec:double-trace} of the present work. We would like
to thank Sumit Das, Avinash Dhar, Rajesh Gopakumar, Shiraz Minwalla and Spenta
Wadia for a careful reading of the manuscript and for
important feedbacks.
We would also like to thank user support team of
Strategic Programs for Innovative Research (SPIRE) Field 5 ``The origin of matter and the universe'', especially Satoru Ueda for helping our numerical calculation in Section \ref{sec:single-trace}.
We also thank Lowell D. Johnson, who is the contributor of the GSL Mathieu function routines, for useful advice about the routines.
The work of T.M. is supported in part by Grant-in-Aid for Scientific Research (No. 24840046) from JSPS.

\appendix
\section{\label{app:droplet} Fermionic formulation of MQM}

In this appendix, we will discuss the fermionic formulation of
MQM. For a general review, see, e.g.  \cite{Ginsparg:1993is,
  Nakayama:2004vk}.\footnote{For specific details of the material
  presented here, see, e.g.  \cite{Sengupta:1990bt, Mandal:1991ua,
    Das:1990kaa, Gross:1990st, Polchinski:1991uq, Dhar:1992rs,
    Dhar:1992hr}.}

\subsection{Single trace model}

Let us first consider the single-trace model \eq{single-trace}.  The
singlet sector of this model (see footnote \ref{ftnt:singlet}) is
described by the following $N$-fermion action and hamiltonian
\begin{align}
S &= \int dt~ \sum_{i=1}^N \left( \frac{\dot \t_i^2}{2} - a \cos \t_i \right),
\nn
H &= \sum_{i=1}^N h(\t_i, \del_{\t_i}), \quad h = - \frac{\hbar^2}2 \del_\t^2 + 
V(\t), \quad V(\t)= a \cos \t, \quad \hbar=1/N  .
\label{particle-single}
\end{align}
The positions $\t_i$ are defined on a circle, and arise from the
eigenvalues of $U= V$ diag$[e^{i\t_i}] V^{-1}$; the fermionic nature
arises from the measure (after integrating out the $V_{ij}$'s) which
vanishes for coincident eigenvalues. The identification of $\hbar$ as
$1/N$ comes from the partition function in \eq{unitary-action}.

A second-quantized version of the above action is written down:
\begin{align}   
S &= \int dt~ d\theta~ \psi^\dagger(\theta,t)\left[-i \hbar \del_t -
  h(\theta, \del_\theta)\right] \psi(\theta,t), 
\quad \int d\t~\psi^\dagger (\t,t) \psi(\t,t) = N.
\label{fermion-single}
\end{align}
It is useful to describe states of the system in terms of expectation
value of the phase space (Wigner) distribution
\begin{align}
\hat u(\t,p,t) = \int d\eta~\psi^\dagger \left(\t+ \frac\eta 2,t\right) \psi \left(\t-
\frac\eta 2, t\right) e^{i p \eta/\hbar}.
\label{wigner}
\end{align}
We define the position space density as 
\begin{align}
\rho(\t,t)= \frac1N\sum_{i=1}^N \delta(\t - \t_i(t))= \frac1N
\psi^\dagger (\t,t) \psi(\t,t) =\frac1N \int \frac{dp}{2\pi \hbar}
\hat u(\t,p,t).
\label{density}
\end{align} 
The normalization condition in \eq{fermion-single} implies that
\begin{align}
\int d\t\ \rho(\t,t) =  \int \frac{d\t\,dp}{2\pi} u(\t,p,t)=1.
\label{normalization}
\end{align} 

\paragraph{Large-$N$ limit} Since $\hbar=1/N$, the large $N$
limit reduces to
\begin{align}
N \to \infty, \quad \hbar \to 0, \quad N\hbar =1.
\label{scaling}
\end{align}
In this limit, the fermion configurations are represented by a
semiclassical Fermi liquid with a smooth phase space density and
position space density; the expectation value $u(\t,p,t)$ is either 0
or 1 depending on whether the phase space cell (of vanishing area
$2\pi\hbar$) is occupied by a fermion or not). The regions with $u=1$
are called droplets of fermi liquid.

\paragraph{Phases:} Classically, due to the potential $V(\t)= a \cos \t$  
in \eq{particle-single} the particles tend to clump around $\t=\pi$;
quantum mechanically however the states of the system are quantized
and Pauli exclusion tends to spread the fermions up. The competing
tendencies dictate the qualitative nature of the ground state: (a) For $a> a_c = \pi^2/64$, the
potential is strong and the eigenvalues are clumped around $\t=\pi$.
Pauli exclusion forces the fermions to occupy $N$ levels, but the
Fermi level is below the top of the potential. This leads to a gap in
$\rho(\t)$ around $\t=0$.  This is called the {\it gapped phase}. (b)
For $a< a_c$ the potential is weak and the eigenvalues are spread out
(non-uniformly). The Fermi level is above the top of the potential and
there are no gaps in $\rho(\t)$. This phase is called a {\it gapless
  phase}.  The phase transition at $a=a_c$ is characterized by the
closing of a gap, when the Fermi level comes up and touches the top of
the potential. This is a third order phase transition and was found in
 \cite{Wadia:1980cp}.

\paragraph{Double scaling:} It is clear from the above that the position
of the fermi level $\mu$, measured from the top of the potential, is
$O(a-a_c)$. It turns out that the singularity of the free energy is
the form $F_{sing}(\mu N) $$ \propto f_0 (\mu N)^2 + f_1 + f_2 (\mu
N)^{-2}+...$. The double scaling limit is defined by $N\to \infty, \mu
\to 0$, such that $\mu N$ is held fixed. Our result about the entropy 
$S \sim O(N)$ is without taking into account such a double scaling.

\paragraph{The hermitian model:} The double scaling limit captures
the universal part of the potential near the top: $\t \approx 0$,
whose essential features can be equally simply reproduced, by
replacing $\cos\t$ by ($1- \t^2/2$) in \eq{particle-single}; dropping
the 1, the potential becomes $V(\t)= -a \t^2/2$, which is equivalent
to the (singlet sector of the) hermitian matrix model
\begin{align}
Z_H = \int DM(t) \ e^{iN S}, \quad S = \int \kern-5pt dt \left[\frac{1}{2} 
\Tr {\dot M}^2   
-V(M) \right], \quad V(M)= - \frac{a}{2} \Tr M^2   .
\label{hermitian}
\end{align}
This model is called the $c=1$ matrix model.\footnote{The periodic
  boundaries $\t= \pm \pi$ need to be supplanted by $\t = \pm \t_m$ to
  make the model well-defined; the choice of $\t_m$ is not
  particularly relevant and does not affect critical/universal
  quantities.} The double-scaled $c=1$ model is mapped to
two-dimensional bosonic string theory (see the review
\cite{Ginsparg:1993is}) by identifying $N\mu= 1/g_{str}$.  A similar
duality to two-dimensional type 0 theories also exists (see the review
\cite{Nakayama:2004vk}).

\paragraph{Classical droplet motion:} Suppose that
the Fermi liquid configuration is that of a single droplet (e.g.  the
Fermi sea or a continuous distortion of it), characterized by
specifying two functions $p_\pm(\t,t)$, which demarcate the
upper/lower boundaries of the droplet at a specific value of $\t$ at
time $t$, i.e
\begin{align}
u(\t,p,t)= \theta(p_+(t)-p) \theta(p-p_-(t)).
\label{quad-profile}
\end{align}
Using this, we get the following formulae for density, momentum
density and energy density:
\begin{align}
\rho(\t,t) &= \int \frac{dp}{2\pi} u(\t,p,t) = \frac1{2\pi}
\left(p_+(\t,t)- p_-(\t,t)\right),
\nn 
\rho(\t,t) {\cal P}(\t,t) &= \int\frac{dp}{2\pi} ~p u(\t,p,t) = 
\frac1{2\pi} (p^2_+(\t,t)-
p^2_-(\t,t))/2 ,
\nn 
H &= \int \frac{d\t~dp}{2\pi} h(p,\t) u(\t,p,t) = \int
d\t \int_{p_-(\t,t)}^{p_+(\t,t)} \frac{dp}{2\pi} (p^2/2 + V(\t)),
\nn 
&= \int {d\t}\ \left[ \frac12 \rho {\cal P}^2 + \frac{\pi^2}6 \rho^3
+ \rho(\t) V(\t)  \right].
\label{collective-single}
\end{align}
Here $V(\t)= a \cos \t$ for the unitary matrix model \eq{single-trace},
\eq{fermion-single}, whereas $V(\t)= - a\t^2/2$ for the hermitian
matrix model \eq{hermitian}.
It turns out that $\rho(\t,t),{\cal P}(\t,t)$ are canonically conjugate
variables, with Poisson bracket
\[
\{\rho(\t,t),{\cal P}(\t',t)\}= - \del_\t \delta(\t - \t')  .
\]
This can be derived from the anticommutation relation of the fermion
field, or from the symplectic $W_\infty$ structure of the
$u(\t,p)$-fields (see \cite{Dhar:1992rs, Dhar:1992hr}). The equation
of motion of the `collective fields' $\rho,{\cal P}$  or $p_\pm$ follow from
the above hamiltonian and Poisson brackets (see comments below
\eq{u-free} for limitations of the collective field description):
\begin{align}
 \dot \rho(\t,t)=  -  \del_\t(\rho(\t,t){\cal P}(\t,t)), \quad
\dot {\cal P}(\t,t)= 
-\del_\t\left[\frac12 {\cal P}^2+ \frac{\pi^2}2 \rho^2 + V(\t)\right].
\label{rho-EOM}
\end{align}
These equations of motion (EOM) can alternatively be derived from that
of the phase space density
\begin{align}
\left(\del_t - \frac{\del h}{\del \t} \del_p + \frac{\del h}{\del p} \del_\t \right)
u(\t,p,t)=0 ,
\label{u-EOM}
\end{align}
which, viewed from the particle-fixed frame, is simply the particle
EOM, and can be regarded as the Euler equation or the dissipationless
Boltzman equation. This viewpoint provides an explicit solution of
the equation of motion:
\begin{align}
u(\t,p,t)= u_0(\t_0(\t,p,t), p_0(\t,p,t)),
\label{u-sol}
\end{align}
 where  $(\t_0, p_0)$ is the unique initial
phase space point at $t=0$ which reaches $(\t,p)$ at time $t$ through 
the dynamical evolution \eq{particle-single},  
and $u_0$ is the phase space density at $t=0$.
 For the hermitian model \eq{hermitian}, $V
= - a \t^2/2$, and \eq{u-sol} becomes
\begin{align}
u(\t,p,t)= u_0\left(\t \cosh (\sqrt a t) -\frac{p}{\sqrt a} 
\sinh (\sqrt a t),-\t \sqrt a \sinh
(\sqrt a t) + p \cosh (\sqrt a t)\right),
\label{u-hermitian}
\end{align} 
which can be easily verified to be a solution of \eq{u-EOM} for
$h=p^2/2-a\t^2/2$. In the special case of $a=0$ (free fermions
with zero potential), we get
\begin{align}
u(\t,p,t)= u_0(\t - p t,p).
\label{u-free}
\end{align}
Note that the collective field equations of motion \eq{rho-EOM}
depend on the assumption of a quadratic profile \eq{quad-profile},
which are clearly violated by `folds' e.g. in  Fig.~\ref{fig:dyn-phase-tr}
and Fig.~\ref{fig-thermalization}. As a result, they are not
valid for long times (see \cite{Dhar:1992rs,Dhar:1992hr}). In our actual analysis where we are specially
interested in long time evolutions, we do not use \eq{rho-EOM}
but use \eq{u-EOM} or equivalently the particle equations of motion
in Appendix \ref{app:details}.

Given the solution \eq{u-sol}, and the relation \eq{density} between
$\rho$ and $u$, we get the following result for $\rho(\t,t)$ at
$N\to \infty$:
\begin{align}
\rho(\t,t)= \int \frac{dp}{2\pi}  u_0(\t_0(\t,p,t), p_0(\t,p,t)).
\label{rho-t-coll}
\end{align}
This can be numerically implemented by sprinkling fermions 
uniformly in the support of the original function $u_0$ and 
let each fermion evolve according to the EOM of \eq{particle-single}.  

\subsection{Double trace model}

Using the fermionization techniques discussed in the previous
subsection, the action \eq{double-trace} can be written as (with
$\hbar=1/N$, as before)
\begin{align}
S &= \int dt\left[ \sum_{i=1}^N \frac{\dot\t_i^2}2 - \frac\xi{N}
\sum_{i,j} \cos(\t_i - \t_j) \right]
\label{particle-double}
\\
&= \int dt~ d\theta~ \psi^\dagger(\theta,t)[-i\hbar
\del_t - h(\theta, \del_\theta)] \psi(\theta,t)- H_{int},\qquad
h(\theta, \del_\theta) = - \hbar^2 \del_\theta^2, 
\nn
H_{int} & = \frac\xi{N} \int dt~d\t~d\t'\left[\psi^\dagger(\t,t)\psi(\t,t)
\cos(\t- \t')\psi^\dagger(\t',t)\psi(\t',t)\right].
\label{fermion-double}
\end{align}
The definition of $u(\t,p,t), \rho(\t,t)$, ${\cal P}(\t,t),
p_\pm(\t,t)$ all remain as before. 
\begin{align}
H&= \int {d\t}\ \left[ \frac12 \rho {\cal P}^2 + \frac{\pi^2}6 \rho^3
\right] + \frac\xi N \int d\t~d\t' \rho(\t) \cos(\t-\t') \rho(\t').
\label{collective-double}
\end{align}
Because of the limitation of the collective field description
in the presence of folds (see comments below \eq{u-free}) for
computation of the time evolution we use the equation of
motion of the phase space density or equivalently the particle
equations of motion as in Appendix \ref{app:details}.

\section{\label{app:GGE}Integrability and GGE}

The single-trace model \eq{particle-single}, \eq{fermion-single} is
clearly integrable, since it consists of $N$ free particles.
Classically one can consider the energy $\ep_i$ of each particle,
$i=1,...,N$, to be an `action' variable. Quantum mechanically, each
linearly independent $N$-fermion state can be specified by 
saying which single-particle energy levels, $\ep_i$, are occupied.  
The quantities
\begin{align}
I_p = \sum_{i=1}^N \ep_i^p,
\label{i-p}
\end{align}
are all constant in time, and commute with each other. For finite $N$,
only $N$ of them, $I_p$ $(p=1,...,N)$ is enough to determine the energy
levels $\ep_i$ (up to some discrete ambiguities which we shall
ignore), hence $I_{N+1}, I_{N+2},...$ are all dependent on the first
$N$ $I_p$'s. In the $N\to \infty$ limit, of course, the charges
\eq{i-p} are all independent. 

We can define the Generalized Gibbs Ensemble (GGE) in terms of the
$N$ conserved charges as follows:
\begin{align}
\varrho_{\rm GGE}= \frac1{Z_{\rm GGE}}\exp[-\sum_p \b_p I_p], \quad
Z_{\rm GGE}=   \Tr \exp[-\sum_p \b_p I_p].
\label{GGE}
\end{align}
Here $\beta_p$ are $N$ chemical potentials. 
For more details, see \cite{Polkovnikov:2010yn}.

In the present discussion, it is convenient to introduce the fermion
occupation numbers $N_m$, $m=0,1,..., \infty$, where $N_m=1$ if the
single particle energy level $\ep_m$ is occupied, and $=0$ if it is
not (note that our fermions are spinless, and that the energy levels
are non-degenerate). These can be defined in terms of the
creation/annihilation operators through the equations 
\begin{align}
\psi(\t) &=  \sum_m  c_m \phi_m(\t), \quad h(\t,\del_\t)\phi_m(\t)
= \ep_m \phi_m(\t),
\label{expansion}\\
N_m &\equiv  c^\dagger_m c_m  .
\label{def-N-m}
\end{align}
In terms of the $N_m$'s, $I_p= \sum_m \ep_m^p N_m$. For $N= \infty$,
in stead of \eq{i-p}, one can use $N_m$'s as the (independent) conserved
charges, which leads to the following definition of the GGE:
\begin{align}
\varrho_{\rm GGE}= \frac1{Z_{\rm GGE}}\exp[-\sum_m \mu_m N_m], \quad
Z_{\rm GGE}=   \Tr \exp[-\sum_m \mu_m N_m].
\label{GGE-N-m}
\end{align}
With this definition, we get, by explicit evaluation (see \cite{Rigol:2007}),
\begin{align}
\ln Z = \sum_m \ln(1 + e^{-\mu_m}), \quad
\vev{N_m}_{\rm GGE}= - \frac{\del}{\del {\mu_m}} \ln Z =  \frac1{e^{\mu_m}+1},
\label{N-m-GGE}
\end{align} 
which gives a remarkably simple Fermi-Dirac distribution which
involves only one chemical potential. The chemical potentials are
now trivially found: 
\begin{align}
\mu_m =  \ln\left( \frac1{\vev{N_m}_{\rm GGE}} -1 \right).
\label{chem-pot}
\end{align} 
The entropy of the GGE is given by
\begin{align}
S_{\rm GGE} = -\Tr \varrho_{\rm GGE} \ln \varrho_{\rm GGE} = -\kern-5pt 
\sum_{m=0}^\infty 
\left[ \vev{N_m}_{\rm GGE} \ln \vev{N_m}_{\rm GGE} +  
(1- \vev{N_m}_{\rm GGE}) \ln (1- \vev{N_m}_{\rm GGE}) \right].
\label{GGE-entropy}
\end{align}
The GGE hypothesis is that, if we start from an appropriate excited
state $| \Psi(0) \ran$, there exists a certain class of observables
$O$ such that, as $t\to \infty$
\begin{align}
\lan \Psi(t) | O | \Psi(t) \ran \to \Tr \left(\varrho_{\rm GGE} O\right),
\label{GGE-hypo}
\end{align}
where the GGE is defined by the charges
\begin{align}
\vev{N_m}_{\rm GGE} = \lan \Psi(0) | N_m | \Psi(0) \ran = \lan \Psi(t) |
N_m | \Psi(t) \ran,
\label{N-m}
\end{align}
or equivalently by the chemical potentials $\mu_m$, determined through
\eq{chem-pot}. 

Note that if the initial state $| \Psi(0) \ran$ is an eigenstate
of the number operators $N_m$, then by \eq{N-m},  $\vev{N_m}=0$
or $1$. The GGE constructed from such an initial state has
vanishing entropy $S_{\rm GGE}=0$, according to \eq{GGE-entropy}.

\subsection{Verification of the GGE hypothesis for
$\rho(\t)$}

We wish to verify \eq{GGE-hypo} where the observable is
the density variable 
\[
O(t) = \rho(\t,t) = \sum_{m,n} c^\dagger_m(t) c_n(t) \phi^*_m(\t)
\phi_n(\t),
\]
where $c_m(t)= c_m \exp[i\ep_m t/\hbar]$, $c^\dagger_m(t)= c^\dagger_m
\exp[-i\ep_m t/\hbar]$. The LHS of \eq{GGE-hypo} becomes, in the
Heisenberg picture,
\begin{align}
\vev{\rho(\t,t)} &=  \sum_{m,n} \vev{c^{\dagger}_m
  c_n}\phi^{*}_m(x) \phi_n(x) \exp[-i(\ep_m- \ep_n)t/\hbar],
\label{rho-t-general}
\end{align}
where $\vev{...} = \lan \Psi(0) | ... | \Psi(0) \ran $ represents
expectation value in the initial state $ | \Psi(0) \ran$.  We will
evaluate this expression in \eq{rho-t} where the initial state is
prepared according to QQD. The large $N$ limit of \eq{rho-t-general}
is given by the semiclassical expression \eq{rho-t-coll}.

To compute the RHS of \eq{GGE-hypo} for $O=\rho(\t)$, first note
that 
\begin{align}
\vev{c^\dagger_m(t) c_n(t)}_{\rm GGE}=  \delta_{mn} \vev{N_m}_{\rm GGE},
\label{GGE-off-diag}
\end{align}
where $\vev{...}_{GGE}= \Tr(\varrho_{\rm GGE}[...])$.
Using this and \eq{expansion}, we get
\begin{align}
\vev{\rho(\t)}_{\rm GGE}= \sum_m \vev{N_m}_{\rm GGE} |\phi_m(\t)|^2.
\label{rho-t-GGE-general}
\end{align}
Note that this is independent of time, as expected. Here $\vev{N_m}_{\rm
  GGE}$ are to be determined from their values in the initial
state  $| \Psi(0) \ran$
\begin{align}
\vev{N_m}_{\rm GGE}= \lan \Psi(0) | N_m | \Psi(0) \ran.
\label{N-m-value}
\end{align}
Thus, 
\begin{align}
\vev{\rho(\t)}_{\rm GGE}= \sum_m 
\lan \Psi(0) | N_m | \Psi(0) \ran |\phi_m(\t)|^2.
\label{rho-GGE-general-value}
\end{align}
Since these uniquely determine the chemical potentials through
\eq{chem-pot}, the GGE is specified completely. 

Verifying the GGE hypothesis amounts to showing that
\eq{rho-t-general} tends to \eq{rho-GGE-general-value} as $t \to
\infty$. In Section \ref{sec:QQ} we will compute both these
expressions, and in Section \ref{sec:GGE-text} present the result in
support of the hypothesis.

\subsection{\label{sec:u-GGE}Phase space density in GGE}

In this section, we will compute $\vev{\hat u(\t,p)}_{GGE}$, where
$\hat u(\t,p)$ is the Wigner phase space density \eq{wigner}. By using
\eq{expansion} and \eq{GGE-off-diag}, it is easy to show that
\begin{align}
\vev{\hat u(\t,p)}_{\rm GGE} &= \sum_m \vev{N_m}_{\rm GGE} u_m(\t,p),
\nn u_m(\t,p) &= \int d\eta~\phi_m^* \left(\t+ \frac\eta 2 \right) \phi_m \left(\t-
\frac\eta 2 \right) e^{i p \eta/\hbar},
\label{u-GGE}
\end{align}
where $ \vev{N_m}_{\rm GGE}$ is given by \eq{N-m-value}. In 
Section \ref{sec:coll-GGE} we evaluate this expression in
the semiclassical limit.

\section{\label{sec:QQ}Quantum quench dynamics (QQD)}

Suppose we consider a sudden change of the parameter $a$ in the
single-particle hamiltonian \eq{particle-single}, from $a_i$ at $t<0$
to $a_f$ at $t\ge 0$. Time evolution works as follows. We consider the
ground state $|F' \ran$ of the system (Fermi sea in this case)
appropriate to the parameter $a_i$ for $t<0$. Quantum quench dynamics
refers to the sudden approximation in which the system evolves for
$t\ge 0$ according to the new parameter $a_f$, from an initial state
equal to the old Fermi sea 
\begin{align}
|\Psi(0) \ran = | F' \ran.
\label{old-sea}
\end{align}  
Clearly the system is integrable after $t=0$ and our discussion in the
previous section applies (we need to identify the $a$ parameter with
$a_f$).

We wish to compute the time-dependence of
\begin{align}
\vev{\rho(\t,t)} \equiv   \lan F' | \rho(\t,t) | F' \ran =
\lan \Psi(t) | \rho(\t) | \Psi(t) \ran,
\label{rho-t-def}
\end{align}
for $t\ge 0$. Here the second expression is in the Heisenberg picture
(in which the state remains the initial state \eq{old-sea}), while the
third expression is in the Schrodinger picture in which $| \Psi(t)
\ran$ is the time evolution of \eq{old-sea} according to the new
Hamiltonian.

To proceed, let us express the old Fermi sea $| F' \ran$ in terms of
eigenstates in terms of the new number operators $N_m$. Suppose we
denote the eigensystem of the old hamiltonian (with $a_i$) as
$\phi'_m, \ep'_m$. The functions $\phi'_m$ can clearly be written in
the new eiganbasis $\phi_n$: thus
\[
\phi'_m = \sum_m B_{mn} \phi_n, ~ \phi_n = \sum_m  \phi'_m B^*_{mn}.
\] 
Just like \eq{expansion}, the Fermi field can be equally well expanded
in terms of $\phi'_m$; thus
\begin{align}
\psi(\t)= \sum_n c_n \phi_n(\t)=  \sum_m c'_m \phi'_m,
\label{both-expansions}
\end{align}
we get
\begin{align}
c'_m = B^*_{mn} c_n, c^{'\dagger}_m =  B_{mn} c^\dagger_n,
~  c_n = c'_m B_{mn}, c^{\dagger}_n =  c^{'\dagger}_m B^*_{mn}.
\label{bogol} 
\end{align}
Using the expression \eq{rho-t-general} we get
\begin{align}
\vev{\rho(\t,t)} 
&=  \sum_{m,n=1}^\infty \sum_{r=1}^N {B^*_{rm} B_{rn}} \phi^{*}_m(x)
\phi_n(x)  \exp[-i(\ep_m- \ep_n)t/\hbar],
\label{rho-t}
\end{align}
where we have used \eq{bogol} and the defining property 
of the old Fermi-sea: 
\begin{align}
\lan F' |  c^{'\dagger}_n c'_k  | F' \ran = N'_{nk},
\quad N'_{nk} = f_n \delta_{nk},
\quad f_n =1~ (n=1,\dots, N), \quad f_n=0~(n>N).
\label{old-off-diag}
\end{align} 
We explore \eq{rho-t} numerically in the text for various values of
$N$. The $N\to \infty$ limit can be computed in the semiclassically
in the droplet picture as in \eq{rho-t-coll}.

\subsection{\label{sec:GGE-QQ}GGE for QQD}

We wish to compute $\vev{\rho(\t)}_{\rm GGE}$ in the context of QQD,
using the general expression \eq{rho-GGE-general-value}, 
for the initial state \eq{old-sea}. 

The expression \eq{N-m-value} becomes
\begin{align}
\vev{N_m}_{\rm GGE}= \lan F' |  c^\dagger_m c_m | F' \ran =
\lan F' |  c^{'\dagger}_n B^*_{nm} c'_k  B_{km} | F' \ran    
= \sum_{n=1}^N |B_{nm}|^2,
\label{N-m-sudden}
\end{align}
where we have used \eq{old-off-diag} and \eq{bogol}. Using this
expression in \eq{rho-t-GGE-general}, we now get
\begin{align}
\vev{\rho(\t,t)}_{\rm GGE}= \sum_m  \sum_{n=1}^N |B_{nm}|^2  |\phi_m(\t)|^2.
\label{rho-t-GGE}
\end{align}

In Section \ref{sec:GGE-text} we verify the
hypothesis \eq{GGE-hypo} for the fermion density $\rho(\t,t)$ (more
precisely, its various moments \eq{moments}), by comparing the large
time behaviour of \eq{rho-t} (or \eq{rho-t-coll}) and \eq{rho-t-GGE}.

\subsection{\label{app:adiabatic}Adiabatic time evolution}

Suppose we have a time-dependent Hamiltonian $H(t)$, e.g.
\eq{fermion-single} with $a= a(t)$-parameter, or \eq{fermion-double}
with $\xi= \xi(t)$-parameter. Suppose the time-dependence starts after
some time $t_0$; we wish to study the time-evolution of a wavefunction
$| \Psi(t) \ran$, which at $t_0$ is an eigenstate: $H(t_0) | \Psi(t_0)
\ran $ = $E_0 | \Psi(t_0) \ran$. QQD is an example in which the
time-variation is in the form a step function at $t=0$.  An adiabatic
time evolution, on the other hand, refers to a sufficiently slow
time-variation such that one can approximate $| \Psi(t) \ran$ as the
instantaneous eigenstate of $H(t)$ with eigenvalue $E(t)$; this notion
is well-defined in the absence of level crossing, which can be ensured
if the spectrum is discrete which sets a finite time scale. In the
examples considered in this paper, we are often interested in the
time-evolution of the ground state, and by an adiabatic expectation
value $v_{adiab}(t)$ we will always mean expectation value in the
instantaneous ground state $v_{inst}(t)$ = $\lan
\Psi_{ground,inst}(t)| O | \Psi_{ground,inst}(t) \ran$.\footnote{At a
  critical point, the spectrum is gapless; there we will simply {\it
    define} $v_{adiab}(t) \equiv v_{inst}(t)$.}

In the context of \eq{fermion-single}, the ground state of a specific
value $a(t_1) \equiv a_1$ is clearly given by, in an obvious notation
where we denote all instantaneous quantities by an additional
subscript $(...)_1$
\begin{align} 
\lan F_1 | c^{\dagger}_{1,m} c_{1,n} | F_1 \ran= \delta_{mn}
f_{1,m}, ~ f_{1,m} = \sum_{n=1}^N \delta_{nm}.
\label{new-sea}
\end{align}
By expanding $\psi(\t)$ in this new basis, we get 
\begin{align}
\vev{\rho(\t)}_{adiab} \equiv \lan F'_0 | \rho(\t) | F'_0 \ran=
\sum_{m-1}^N |\phi'_m(\t)|^2.
\label{adiabatic-single}
\end{align}
The time dependence of the adiabatic quantity is reflected in the
appearance of the instantaneous eigenfunctions.

\section{\label{app:free}An analytic calculation of relaxation} 

In this section, we will consider time evolution in the single-trace
model \eq{single-trace} with $a=0$.  (This model is also equivalent to
$\xi=0$ in (\ref{double-trace}).)  In this case, the ground state is
given by $p_{\pm}(\t)=\pm 1/2$ (see \eq{collective-single} for
notation).\footnote{Note that this corresponds to $\rho(\t)=
  \frac1{2\pi}, {\cal P}(\t)=0$, which obviously satisfies
  \eq{rho-EOM} for $V=0$.} Suppose we perturb the ground state
configuration at $t=0$ by a sinusoidal deformation
\begin{align} 
p_+(\theta,t=0)=\frac{1}{2}+b \cos(n \theta), \quad (|b|\le 1/2),
\qquad p_-(\theta,t=0)=-p_+(\theta,t=0),
\label{ini-perturbation}
\end{align} 
where $b$ is the amplitude of the perturbation (see Fig.~\ref{fig-thermalization}). We wish to evaluate the time evolution of
this perturbation using the action \eq{fermion-single} with $a=0$.

\begin{figure}
\begin{center}
\includegraphics[scale=.55]{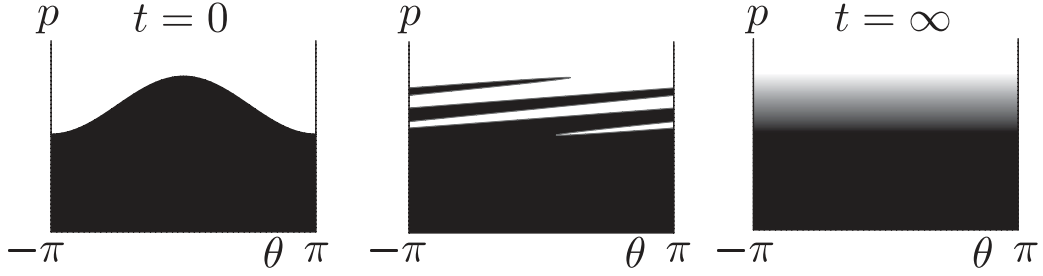}
\caption{\footnotesize Equilibration of the phase space density in the
  $V(U)=0$ case. The perturbation (\ref{ini-perturbation}) with $n=1$
  at $t=0$ (the left figure) evolves to the right configuration.  }
\label{fig-thermalization}
\end{center}
\end{figure}

We will evolve the configuration as in \eq{u-free}, which essentially
says that $u_t(\t_0+p_0 t, p_0) = u_0(\t_0,p_0)$; in other words, to
get any point of the evolved Fermi surface at time $t$, we should
simply evolve points on the initial Fermi surface according to the
equation of motion of (\ref{particle-single}) with $a=0$, which is given by
\begin{align} 
\dot p=0 , \quad \dot \theta = p.
\label{eom-thermalization}
\end{align}
In order to solve this equation for points on the Fermi surface, it is
convenient to employ a parameter $s\in [-\pi,\pi]$ and rewrite the
initial configuration (\ref{ini-perturbation}) as \footnote{From now,
  we omit $p_-$, since we can derive it trivially from $p_+$.}
\begin{align} 
p_+(s,t=0)=\frac{1}{2}+b \cos(n s), \quad \theta(s,t=0)=s.
\end{align} 
Then from (\ref{eom-thermalization}) we obtain the Fermi surface at
time $t$ as
\begin{align} 
p_+(s,t)=\frac{1}{2}+b \cos(n s), \quad \theta(s,t)=s + \left(\frac{1}{2}+b \cos(ns) \right)t.
\label{soln-thermalization}
\end{align} 
This solution with $n=1$ is plotted in the middle of Fig.~\ref{fig-thermalization}; the `folds' appear because of the
stretching caused by different speeds of fermions at different heights
from the $\t$-axis. As time progresses, more or more `folds' keep
appearing.

In the perturbation (\ref{ini-perturbation}), only $\rho_n$ is excited
and other $\rho_m$ are zero.  We evaluate how these modes behave as
$t$ increases.  From the solution (\ref{soln-thermalization}), we
calculate the $m$-th moment as\footnote{The expression in the first
  line in this equation is crude, since the droplet has many folds at a
  late time as shown in the center of Fig.~\ref{fig-thermalization}
  and we need to subtract the white region below $p_+$.  However after
  the change of the variable from $\theta$ to $s$, this subtraction is
  automatically involved.}
\begin{align} 
\rho_m(t)=&\int_{-\pi}^\pi \frac{ d\theta}{2\pi} \cos(m \theta)\left(
p_+-p_- \right) = \int_{-\pi}^\pi \frac{ d s}{2 \pi} \frac{ d
  \theta(s,t) }{d s} \cos(m \theta(s,t))\left( p_+-p_- \right)
\nonumber \\
=&  \int_{-\pi}^\pi  \frac{ d s}{ 2\pi}
\left( 
1-nbt \sin \left( ns \right) 
\right) 
  \cos\left(ms+ mt\left( \frac{1}{2}+b\cos(ns)\right)   \right) 
\left( \frac{1}{2}+b \cos(n s)\right) \nonumber \\
=& \frac{2}{\pi} \int_0^\pi ds  \cos(m s) \cos\left( mt\left( 
\frac{1}{2}+b\cos(ns)\right)   \right) 
\left( \frac{1}{2}+b \cos(n s)\right) \nonumber \\
&+\frac{2}{\pi}\int_{0}^\pi ds~ nbt \sin (ns) \sin(m s) \sin\left( mt\left(
\frac{1}{2}+b\cos(ns)  \right) \right)  \left( \frac{1}{2}+b \cos(n s)\right).
\label{late-moment}
\end{align} 
These integrals will be given by the Bessel function $J_k$.
For example, if $n=1$, the integral
(\ref{late-moment}) with $m=1$ and $2$ are evaluated as
\begin{align} 
\rho_1(t)=\frac{2 }{t} J_1(bt) \cos\left( \frac{t}{2}\right), \qquad 
\rho_2(t)=-\frac{1 }{t} J_2(2bt) \sin \left( t \right).
\end{align} 
For large $t$, they decay with $t^{-3/2}$ as 
\begin{align} 
\rho_1(t) \to \frac{2}{t^{3/2}} \sqrt{\frac{2 }{ \pi b} } \cos \left(  
bt-\frac{3\pi}{4} \right)   \cos\left( \frac{t}{4\pi}\right) , \quad (t 
\to \infty),
\end{align} 
and 
\begin{align} 
\rho_2(t) \to  \frac{1}{t^{3/2}} \frac{1}{ \sqrt{\pi b }} \cos\left( 2bt-
\frac{\pi}{4} \right)  \sin\left( t\right) , \quad (t \to \infty).
\end{align} 
Other $\rho_m(t)$ also exhibits a similar decay. In Section
\ref{sec:cascade} we will compare the time-evolution of various modes
and find a picture of an energy cascade in which lower
frequency modes die out before the higher frequency modes,
causing a transfer of energy from higher to lower frequencies,
in a manner similar to energy cascades in turbulence. 

\subsection{Asymptotic  and GGE form of the Fermi surface}
\label{sec:coll-GGE}

As we mentioned below \eq{soln-thermalization}, the asymptotic form of
the droplet consists of an infinite number of `folds' which
eventually densely fill an entire band. When suitably coarse-grained
(the `grain' size can be finer and finer as time progresses), this
describes a thinned out Fermi liquid, as schematically plotted on the
right of Fig.~\ref{fig-thermalization}.  

Since we have already verified the GGE hypothesis of our system for
the density variable, let us describe the above `thinned out'
Fermi liquid in terms of the GGE. We begin with  the GGE value
\eq{u-GGE} of the phase space density. The single-particle Hilbert
space can be labelled by the momentum wavefunctions
\[
\phi_m(\t)= \frac1{\sqrt{2\pi}} e^{im \t}, m=0,\pm 1, \pm 2,....,
\] 
which trivially gives $u_m(\t,p)= \delta(m- p/\hbar)$, and hence by
\eq{u-GGE} (keeping the notation $\vev{...}_{\rm GGE}$ implicit)
\begin{align}
u(\t,p)= \sum_m N_m \delta(m- p/\hbar) 
= N_{p/\hbar} \equiv \frac{\tilde\rho(p)}{2\pi}.
\label{N-m-coll}
\end{align}
We can identify $\bar\rho(p)$ as the momentum-space density
\begin{align}
\tilde\rho(p) \equiv \int d\t u(\t,p).
\label{mom-density}
\end{align}  
We can compute $N_m$'s from a given initial quantum state by using
\eq{N-m-value}. However, the calculation is much better done
semiclassically, when the initial state is given in terms of a
classical droplet configuration \eq{ini-perturbation}.  Note, first of
all, that the momentum-space density \eq{mom-density} in the present
case is a constant of motion. This is easy to see since $h=p^2/2$ and
all fermions have a constant momentum so the momentum density remains
constant.\footnote{It can also be seen from \eq{u-EOM} and
  \eq{mom-density}.}  We can, therefore, compute the momentum 
density from the initial configuration \eq{ini-perturbation}. Note
that from \eq{ini-perturbation} and \eq{quad-profile} we can
find out the initial value of $u(\t,p)$ and hence $\bar \rho(p)$
by using \eq{mom-density}. The result of this calculation is
\footnote{Geometrically the RHS of \eq{mom-density} is given 
by $\t_+(p) - \t_-(p)$ where $\t_\pm(p)$ are the counterparts
of $p_\pm(\t)$ in \eq{quad-profile}.} 
\begin{align}
u(\t,p)= \frac1{2\pi} \tilde \rho(p)= 
\begin{cases}
 0 & |p| >\frac12 + b, \\
 1  & |p| < \frac12 - b, \\
 \frac1\pi \cos ^{-1}\left(\frac 1b\left(p-\frac{1}{2}\right)\right) &
   \text{otherwise},
\end{cases}
\label{u-val-coll}
\end{align} 
which gives us a quantitative representation of a `thinned out'
Fermi liquid on the right panel of Fig.~\ref{fig-thermalization}.

\subsection{Calculation of Entropy} 

We will now calculate the entropy
$S_{\rm GGE}$. As we
show in Appendix \ref{sec:coll-GGE}, in the large $N$ droplet
description, the $\vev{N_m}$'s are given by \eq{N-m-coll}.  Combining with
\eq{u-val-coll}, and \eq{GGE-entropy}, we get
\begin{align}
S &=  -\kern-5pt \sum_{m=0}^\infty 
\left[ \vev{N_m}_{\rm GGE} \ln \vev{N_m}_{\rm GGE} +  
(1- \vev{N_m}_{\rm GGE}) \ln (1- \vev{N_m}_{\rm GGE}) \right]
\nn
&= -  2 \int_0^\infty \frac{dp}{\hbar} \left[ u(p) \ln u(p) +  
(1- u(p)) \ln (1- u(p)) \right]
= 2.36\ b\ N
\label{analytic-S}
\end{align}
In the first equality, we have inserted a factor of 2 since
the negative $p$-axis contributes the same amount. Also we
have used $m= p/\hbar = N p$ from \eq{N-m-coll}. In the last
step, note that only $\frac12 - b < p < \frac12 + b$
contributes to the integral, for which we use the value of   
$u(\t,p)\equiv u(p)$ in the third line of \eq{u-val-coll}. Note that
$S$ is exactly linear in $N$ (as well as in the amplitude $b$
of the initial perturbation \eq{ini-density}).

\section{\label{app:details} Details of our numerical analysis}

 We explain the details of our computations in Section \ref{single-trace} and \ref{sec:double-trace}.
\paragraph{Exact computation}
We use an exact computation for the single trace matrix model (\ref{ftnt:singlet}) at finite $N$.
In this case, the eigen function $\phi_m(\theta) $ is given by the Mathieu function.
We numerically evaluate this function by using GCC with the GNU Scientific Library.

\paragraph{Semiclassical computation}
For  the single trace matrix model (\ref{single-trace}) at large-$N$ and the double trace matrix model (\ref{double-trace}), we use a semiclassical approximation.
At the strict large-$N$, the dynamics of the matrix models are described by the droplet $u(\theta,p,t)$ in the phase space.
We discretize the support of the initial droplet by $\hat N$ uniformly
distributed points (we take $\hat{N}$ to be $O(10^4)$.)
Each of these points is then
evolved according to the classical equation of motion 
\begin{align}
\dot \t_i = p_i, \qquad \dot p_i &= a \sin \t_i , 
\label{particle-single-EOM}
\end{align}
for the single trace model (\ref{single-trace}), and 
according to the classical equation
of motion
\begin{align}
\dot \t_i = p_i, \qquad\dot p_i &= - \xi \sum_j \sin(\t_i - \t_j), ~~
\label{particle-double-EOM}
\end{align}
for the double trace model (\ref{double-trace}).  Here $\theta_i$ and
$p_i$ ($i=1,\cdots , \hat{N}$) are the points of the discretized
droplet. Evolving these points in this fashion is equivalent
to a discretization of the Euler equation \eq{u-EOM}.  
By using the Runge-Kutta method, we solve these equations
with a given initial condition\footnote{In the case of the single
  trace model, the motion of the surface of the droplet is sufficient
  to determine the dynamics of the model, since the particles are
  free.  In addition, analytic solutions of the equation of motion
  (\ref{particle-single-EOM}) are given by elliptic functions, and can
  be used to significantly improve the calculation of droplet motion.
  Indeed we use this idea to plot Fig.~\ref{fig:dyn-phase-tr} in the
  hermitian matrix model (\ref{hermitian}).  }.  Note that the above
use of the classical equations means that we are in the strict
large-$N$ limit.  As a result, the number $\hat N$, which is simply a
measure of discretization of the classical phase space, is not related
to $N$ in the original matrix model (while $1/N$ measures 
quantum effects, $1/\hat N$ measures some discretization error
of classical equations).

\end{document}